\documentstyle[prb,aps,psfig]{revtex}

\begin{document}

\title{ Quantum dynamics of charged and neutral magnetic solitons: \\
 Spin-charge separation in the one-dimensional Hubbard model}

\author{Mona Berciu and Sajeev John}

\address{ Department of Physics,University of Toronto, 60 St. George
Street, Toronto, Ontario, M5S 1A7, Canada}

\date{\today}

\maketitle

\pacs{} {Nos. 71.27.+a, 71.10.Fd,  74.20.Mn }

\begin{abstract}
We demonstrate that the Configuration Interaction (CI) Approximation
recaptures essential features of the exact (Bethe Ansatz) solution to
the 1D Hubbard model. As such, it provides valuable route for
describing effects which go beyond mean-field theory for strongly
correlated electron systems in higher dimensions.  The CI method
systematically describes fluctuation and quantum tunneling corrections
to the Hartree-Fock Approximation (HFA). HFA predicts that doping a
half-filled Hubbard chain leads to the appearance of charged
spin-polarons or charged domain-wall solitons in the antiferromagnetic
(AFM) background. The CI method, on the other hand, describes the
quantum dynamics of these charged magnetic solitons and quantum
tunneling effects between various mean-field configurations.  In this
paper, we test the accuracy of the CI method against the exact
solution of the one-dimensional Hubbard model. We find remarkable
agreement between the energy of the mobile charged bosonic domain-wall
(as given by the CI method) and the exact energy of the doping hole
(as given by the Bethe Ansatz) for the entire $U/t$ range.  The CI
method also leads to a clear demonstration of the spin-charge
separation in 1D. Addition of one doping hole to the half-filled
antiferromagnetic chain results in the appearance of two different
carriers: a charged bosonic domain-wall (which carries the charge but
no spin) and a neutral spin-1/2 domain wall (which carries the spin
but no charge).
\end{abstract}

\narrowtext \twocolumn

\section{Introduction}

The two-dimensional (2D) Hubbard model (or a modified version of it)
is widely believed to describe the essential physics of the strongly
correlated electrons in high temperature superconducting
cuprates. \cite{PWscience} Unlike the 1D Hubbard model, an exact
solution of the 2D problem is not known, and one must resort to
approximations.  The validity of such approximations can be tested on
the 1D Hubbard model, which is exactly solvable\cite{wu} using the
Bethe Ansatz.\cite{Bethe} It is highly desirable to develop an
approximation which recaptures essential physical features of the
exact Bethe Ansatz solution and which, at the same time, can be
applied to higher dimensional systems. In this paper, we show that the
Configuration Interaction (CI) Method \cite{CIpaper} is such an
approximation.  We demonstrate that the CI method recaptures essential
quantum tunneling effects that go beyond mean-field theory and that
lead to spin-charge separation in 1D.  The predictions of the CI
method in the 2D case as well as a comparison to cuprate
superconductors are described in detail elsewhere.\cite{next}

We consider the generalized Hubbard Hamiltonian:
\begin{equation}
\label{2.0}
{\cal H}=-t\sum_{\langle i,j \rangle \atop \alpha\beta}^{} \left (
c^{\dagger}_{i\alpha} T^{ij}_{\alpha\beta} c_{j\beta} + h.c. \right) +
U \sum_{i}^{} c^{\dagger}_{i\uparrow} c_{i\uparrow}
c^{\dagger}_{i\downarrow} c_{i\downarrow}
\end{equation}
Here, $c^{\dagger}_{i\alpha}$ is the creation operator for an electron
of spin $\alpha$ at site $i$, and the notation $\langle i,j \rangle$
means that the sum is restricted to nearest-neighbors sites $i$ and
$j$.  The parameters of the problem are the hopping matrix $t$, the
on-site interaction matrix $U$, and the SU(2) matrices
$T^{ij}_{\alpha\beta}$, which describe phase factors and internal spin
rotations acquired by the electron as it hops between sites. For the
1D Hubbard Hamiltonian which we consider in this paper we simply set
$T^{ij}_{\alpha\beta} =\delta_{\alpha\beta}$.  However, for the 2D
system, it has been shown that a non-trivial choice of the matrices
$T^{ij}_{\alpha\beta}$, which describes a $2\pi$ (internal) spin
rotation of the electron as it encircles an elementary plaquette of
the square lattice, may be essential to the cuprate
physics.\cite{1,MB3} For this reason, we develop the CI formalism for
the more general Hamiltonian (\ref{2.0}).

In addition to providing a test to gauge the accuracy of the
Configuration Interaction Method, the 1D results may be directly
relevant to certain high T$_c$ cuprate superconductors.
YBa$_2$Cu$_3$O$_7$ and its close relatives have quasi-one-dimensional
CuO chain structures.  Experiments measuring the dc resistivity
\cite{dc}, the infrared and optical conductivity \cite{ifr} and the
penetration depth in untwinned crystals \cite{unt} and ceramics
\cite{cer} have revealed large anisotropies between the a-direction
(perpendicular to chains) and the b-direction (parallel to chains).
These results suggest that substantial currents are carried along the
chains in both the normal and superconducting state. The source of
superconducting condensate on the chains has not yet been elucidated.
	
We begin Section 2 with a brief review of the Static Hartree-Fock
Approximation (HFA).  The HFA leads to a mean-field ground-state with
properties which are in disagreement with those of the exact
ground-state. For the half-filled (undoped) chain, HFA predicts a
degenerate ground-state with long-range antiferromagnetic (AFM) order
and with staggered magnetic moments aligned along some arbitrary
direction. Even in the presence of a weak external interaction which
creates an easy axis for spin orientation, the HF ground state is
degenerate with the related mean-field in which all the spins have
been flipped ($S \rightarrow -S$).  On the other hand, the true
ground-state of the half-filled system is non-degenerate and has no
long-range order, despite the presence of strong AFM
correlations. When holes are introduced into the chain, the HFA leads
to the creation of static magnetic spin-polaron and domain-wall
solitons which trap the hole. Since these solitons are motionless in
the static HFA, this mean field solution breaks the translational
invariance of the original Hamiltonian. The CI method facilitates the
restoration of both of the above symmetries in accord with the exact
Bethe ansatz ground state.

 The essence of the CI method is to use a linear combination of HF
wavefunctions in order to restore the various broken symmetries of the
mean-field theory. In a doped system, for instance, the CI
wavefunction is chosen to be a linear combination of HF wavefunctions
describing the charged soliton centered at various sites. Besides
restoring the translational symmetry, such a wavefunction also
effectively takes into account the quantum dynamics of the charged
soliton along the chain. This motion represents a large amplitude
tunneling event between a given AFM mean field and the alternative AFM
mean-field obtained by the operation ($S \rightarrow -S$). Moreover,
the propagating soliton lowers its quantum zero point energy
considerably, relative to the static (HFA) soliton.

In Section 3 we briefly review the exact Bethe Ansatz (BA) solution
and the ground-state energies of the half-filled and doped chains.
The comparison between the Bethe Ansatz, Hartree-Fock and
Configuration Interaction solutions is presented in Section 4. For
both the undoped chain, and the chain with one doping hole, we show
that through the nucleation of mobile quantum solitons, the CI method
provides a much better description than the HFA. We identify the
charge carrier of the Hubbard chain to be a charged bosonic
domain-wall, and we show that its energy, in the CI method, is in
excellent agreement with the BA result. The CI method leads to a
simple, physical interpretation of spin-charge separation. Adding a
single hole to a chain leads to the creation of two magnetic
solitons. One soliton is the bosonic charged domain wall (which
carries charge, but no spin), while the other soliton is the neutral
spin-1/2 domain wall (which carries spin, but no charge). The energy
of these excitations are in good agreement with the BA results.
Finally, Section 5 contains discussion of the results and conclusions.

\section{\label{sec2} Approximations: Hartree-Fock and 
Configuration Interaction Method}

\subsection{The Static Hartree-Fock Approximation}

One of the most widely used approximations for the many-electron
problem is the Static Hartree-Fock Approximation (HFA).  In this
approximation the many-body problem is reduced to one-electron
problems in which each electron moves in a self-consistent manner
depending on the mean-field potential of the other electrons in the
system.  While this method is insufficient, by itself, to capture all
of the physics of low dimensional electronic systems with strong
correlations, it provides a valuable starting point from which
essential fluctuation corrections can be included. In particular, we
use the Hartree-Fock method to establish the electronic structure and
the static energies of various magnetic soliton structures. In the
more general Configuration Interaction (CI) variational wavefunction,
the solitons acquire quantum dynamics and describe large amplitude
tunneling and fluctuation effects that go beyond mean field theory.

In the HF approximation, the many-body wavefunction $| \Psi\rangle$ is
decomposed into a Slater determinant of effective one-electron
orbitals. The one-electron orbitals are found from the condition that
the total energy of the system is minimized
\begin{equation}
\label{2.1}
\delta { \langle \Psi| {\cal H}| \Psi\rangle \over \langle\Psi |
\Psi\rangle} =0
\end{equation}

In order to approximate the ground state of the Hamiltonian (1), we
consider a Slater determinant trial-wavefunction of the form
\begin{equation}
\label{2.3}
|\Psi\rangle=\prod_{p=1}^{N_e} a^{\dagger}_p |0\rangle,
\end{equation}
where $|0\rangle$ is the vacuum state, $N_e$ is the total number of
electrons in the system and the one-electron states are given by
\begin{equation}
\label{2.4}
a^{\dagger}_n=\sum_{i \sigma}^{} \phi_n(i,\sigma)
c^{\dagger}_{i\sigma}
\end{equation}
Here, the one-particle wave-functions $\phi_n(i,\sigma)$ form a
complete and orthonormal system.

Using the wavefunction (\ref{2.3}) in equation (\ref{2.1}), and
minimizing with respect to the one-particle wavefunctions
$\phi_n(i,\sigma)$, we obtain the Hartree-Fock eigen-equations:
\parbox{80mm}{
$$ E_n \phi_n(i,\alpha)= -t \sum_{j \in V_i, \beta}
T^{ij}_{\alpha\beta} \phi_n(j,\beta)$$
\begin{equation} 
\label{2.5}
+U \sum_{\beta}^{} \left( {1 \over 2}\delta_{\alpha\beta} Q(i)
-\vec{\sigma}_{\alpha\beta} \vec{S}(i)\right) \phi_n(i,\beta)
\end{equation}}

where $(\sigma_x, \sigma_y, \sigma_z)$ are the Pauli spin matrices and
the charge density,
\begin{equation}
\label{2.6}
Q(i)=\langle\Psi|c^{\dagger}_{i\alpha}c_{i\alpha}|\Psi\rangle=
\sum_{p=1}^{N_e} |\phi_p(i,\alpha)|^2,
\end{equation}
and the spin density,
\begin{equation}
\label{2.7}
\vec{S}(i)=\langle\Psi|c^{\dagger}_{i\alpha} {
\vec{\sigma}_{\alpha\beta} \over 2} c_{i\beta}|\Psi\rangle = \sum_{p=1
}^{N_e} \phi^*_p(i,\alpha){ \vec{\sigma}_{\alpha\beta} \over 2}
\phi_p(i,\beta),
\end{equation}
 must be computed self-consistently.  The notation $j \in V_i$
appearing in (\ref{2.5}) means that the sum is performed over the
sites $j$ which are nearest-neighbors of the site $i$. The
self-consistent Hartree-Fock equations (\ref{2.5},\ref{2.6},\ref{2.7})
must be satisfied by the occupied orbitals $p=1\dots N_e$, but can
also be used to compute the empty (hole) orbitals.

The ground-state energy of the system in the HFA is given by
\begin{equation}
\label{2.8}
E_{GS}=\langle \Psi| {\cal H} |\Psi\rangle= \sum_{p=1}^{N_e} E_p - U
\sum_{i}^{}\left({1\over 4} Q(i)^2- \vec{S}(i)^2\right)
\end{equation}
where the single particle energies are obtained from (\ref{2.5}).

The approximation scheme described so far is called the Unrestricted
Hartree-Fock Approximation, because we did not impose constraints on
the wavefunction $| \Psi\rangle$ which would require it to be an
eigenfunction of various symmetry operations which commute with the
Hamiltonian (\ref{2.0}). If these symmetries are enforced, the method
is called the Restricted Hartree-Fock Approximation.  We use the
Unrestricted HFA since it leads to lower energies. The breaking of
symmetries in our case implies that electronic correlations are more
effectively taken into account. \cite{Fulde} The restoration of these
symmetries is deferred until the CI wavefunction is introduced.
 
In the undoped (half-filled) case, the self-consistent Hartree-Fock
equations can be solved analytically for the infinite system, using
plane-wave one-particle wave-functions.  In the unrestricted
Hartree-Fock approach, doping the system leads to the appearance of
inhomogeneous solutions, which break the translational invariance.  In
this case, we solve the unrestricted self-consistent Hartree-Fock
equations numerically on a finite chain.  Starting with an initial
spin and charge distribution $\vec{S}(i)$ and $Q(i)$, we numerically
solve the eigenproblem (\ref{2.5}) and find the HF eigenenergies $E_n$
and wavefunctions $\phi_n(i,\alpha)$. These are used in
Eqs. (\ref{2.6}) and (\ref{2.7}) to calculate the new spin and charge
distribution, and the procedure is repeated until self-consistency is
reached.  Numerically, we define self-consistency by the condition
that the largest variation of any of the charge or spin components on
any of the sites of the lattice is less that $10^{-9}$ between
successive iterations.

\subsection{Configuration Interaction  Method}

  The basic idea of the Configuration Interaction (CI) method is that
the ground-state wavefunction, for a system with $N_e$ electrons, is
not just a $N_e\times N_e$ Slater determinant (as in the HFA), but a
judiciously chosen linear combination of such Slater
determinants. \cite{CIpaper} Given the fact that the set of all
possible Slater determinants (with all possible occupation numbers)
generated from a complete set of one-electron orbitals constitute a
complete basis of the $N_e$-particle Hilbert space, our aim is to pick
out a subset of Slater determinants which captures the essential
physics of the exact solution.

Consider the CI ground-state wavefunction given by
\begin{equation}
\label{3.1}
|\Psi\rangle= \sum_{i=1}^{N}\alpha_i|\Psi_i\rangle
\end{equation}
where each $|\Psi_i\rangle$ is a distinct $N_e\times N_e$ Slater
determinant and the coefficients $\alpha_i$ are chosen to satisfy the
minimization principle:
\begin{equation}
\label{3.2}
{\delta \over \delta \alpha_i } \left( { \langle\Psi|{\cal
H}|\Psi\rangle \over \langle\Psi|\Psi\rangle} \right) =0\hspace{20mm}
i=1,N
\end{equation}
This leads to the system of CI equations
\begin{equation}
\label{3.3}
\sum_{j=1}^{N}{\cal H}_{ij} \alpha_j=E \sum_{j=1}^{N} {\cal O}_{ij}
\alpha_j
\hspace{20mm} i=1,N
\end{equation}
where $ E= \langle\Psi|{\cal H}|\Psi\rangle / \langle\Psi|\Psi\rangle$
is the energy of the system in the $|\Psi\rangle$ state , ${\cal
H}_{ij}= \langle \Psi_i| {\cal H}|\Psi_j\rangle$ are the matrix
elements of the Hamiltonian in the basis of Slater determinants
$\{|\Psi_i\rangle, i=1,N\}$, and ${\cal
O}_{ij}=\langle\Psi_i|\Psi_j\rangle$ are the overlap matrix elements
of the Slater determinants (which are not necessarily orthogonal). The
CI solution is easily found by solving the linear system of equations
(\ref{3.3}), once the basis of Slater determinants $\{|\Psi_i\rangle,
i=1,N\}$ is chosen.  If we denote by $\phi_p^{(n)}(i,\sigma)$ the
$p=1,...,N_e$ one-electron occupied orbitals of the Slater determinant
$|\Psi_n\rangle$, these matrix elements are given by:
\begin{equation}
\label{3.3b1}
{\cal O}_{nm}= \left|
\begin{array}[c]{ccc}
\beta_{1,1}^{nm} & ... & \beta_{1,N_e}^{nm} \\ \vdots & & \vdots\\
\beta_{N_e,1}^{nm} &... &\beta_{N_e,N_e}^{nm} \\
\end{array}
\right|
\end{equation}
The matrix elements of the Hamiltonian (\ref{2.0}) can be written as:
\begin{equation}
\label{3.3b2}
{\cal H}_{nm}= -t \cdot {\cal T}_{nm}+ U \sum_{i}^{}{\cal V}_{nm}(i)
\end{equation}
where the expectation values of the hopping and on-site interaction
terms are:
$$
{\cal T}_{nm}=\sum_{p=1}^{N} \left|
\begin{array}[c]{ccccc}
\beta_{1,1}^{nm}&...&t_{1,p}^{nm}&...&\beta_{1,N_e}^{nm}\\ \vdots &
&\vdots & & \vdots\\
\beta_{N_e,1}^{nm}&...&t_{N_e,p}^{nm}&...&\beta_{N_e,N_e}^{nm} \\
\end{array}
\right|
$$
and
$$
{\cal V}_{nm}(i)\!\!=\!\!\sum_{p_1 \neq p_2}^{} \!  \left|
\begin{array}[c]{ccccccc}
\beta_{1,1}^{nm} \!\!&...\!\!&u_{1,p_1}^{nm}(i)\!\!& ...
\!\!&d_{1,p_2}^{nm}(i)\!\!&...\!\! &\beta_{1,N_e}^{nm} \\ \vdots & &
\vdots & & \vdots & & \vdots\\
\beta_{N_e,1}^{nm}\!\!&...\!\!&u_{N_e,p_1}^{nm}(i)\!\!& ...
\!\!&d_{N_e,p_2}^{nm}(i)\!\!&...\!\! & \beta_{N_e,N_e}^{nm}\\
\end{array}
\right|
$$
Here,
$$
\beta^{nm}_{ph}= \sum_{{i\sigma}}^{}
\phi^{(n)*}_{h}(i,\sigma)\phi^{(m)}_{p}(i,\sigma)
$$
$$
t^{nm}_{p_1,p_2}= \sum_{\langle i, j \rangle
\atop\alpha\beta}^{}\left( \phi^{(n)*}_{p_1}(i,\alpha)
T^{ij}_{\alpha\beta} \phi^{(m)}_{p_2}(j,\beta) + h.c. \right)
$$
$$
u^{nm}_{p_1,p_2}(i)=\phi^{(n)*}_{p_2}(i\uparrow)\phi^{(m)}_{p_1}(i\uparrow)
$$
$$
d^{nm}_{p_1,p_2}(i)=\phi^{(n)*}_{p_2}(i\downarrow)
\phi^{(m)}_{p_1}(i\downarrow)
$$

We now consider the specific choice of the Slater determinant basis
$\{|\Psi_i\rangle, i=1,N\}$.  Strictly speaking, one may choose an
optimized basis of Slater determinants from the general variational
principle:
\begin{equation}
\label{3.4}
{\delta \over \delta \phi^{(n)}_p(i,\sigma) } \left( {
 \langle\Psi|{\cal H}|\Psi\rangle \over \langle\Psi|\Psi\rangle}
 \right) =0\hspace{5mm} n=1,N; p=1,N_e.
\end{equation}
However, implementation of this full trial-function minimization
scheme (also known as a multi-reference self-consistent mean-field
approach \cite{Fulde}) is numerically cumbersome even for medium-sized
systems. Instead, we select the Slater determinant basis
$\{|\Psi_i\rangle, i=1,N\}$ from the set of broken symmetry,
Unrestricted Hartree-Fock wavefunctions (\ref{2.3}), their symmetry
related partners and their excitations. Clearly, (\ref{2.3}) satisfies
(\ref{3.4}) by itself, provided that the $\alpha$ coefficients
corresponding to the other Slater determinants in Eq.(\ref{3.1}) are
set to zero (see Eq. (\ref{2.1})).  Since this Unrestricted HF
wavefunction is not translationally invariant (the doping hole is
always localized somewhere along the chain), we can restore the
translational invariance of the CI ground-state wavefunction by also
including in the basis of Slater determinants all the possible lattice
translations of this Unrestricted HF wavefunction. In 2D, we must also
include all the possible non-trivial rotations.

Clearly, all the translated HF Slater determinants lead to the same HF
ground-state energy $\langle \Psi_n| {\cal H} | \Psi_n\rangle =
E_{GS}$ as defined by Eq. (\ref{2.8}). The CI method lifts the
degeneracy between states with the hole-induced configuration
localized at different sites, thereby restoring translational
invariance. We may identify the lowering in the total energy due to
the lifting of this degeneracy as quantum mechanical kinetic energy of
deconfinement which the doping-induced configuration saves through
hopping along the chain. In addition, quantum fluctuations in the
internal structure of a magnetic soliton can be incorporated by
including the lowest order excited state configurations of the static
Hartree-Fock energy spectrum. Such wavefunctions are given by
$a^{\dagger}_p a_h |\Psi\rangle$, where $p > N_e$ labels an excited
particle state and $h \le N_e$ labels the hole which is left behind
(see Eq. (\ref{2.3})). Once again, all possible translations of this
``excited'' configurations must be included in the full CI
wavefunction. These additions can describe changes in the ``shape'' of
the soliton as it undergoes quantum mechanical motion through the
crystal.

\section{\label{sec3}Exact 
solution of the 1D Hubbard Model: the Bethe Ansatz}

Before reporting the results obtained in the HF and CI approximation
for the 1D Hubbard model, we briefly describe the exact Bethe Ansatz
solution of this problem, \cite{wu} for comparison purposes.  Consider
an $N$-site chain with $N_e$ electrons of which $M$ have
spin-down. Here, $N_e \le N$ and $M \le N_e/2$.\cite{wu} Any
wavefunction satisfying the many-body Schr\"odinger equation ${\cal
H}|\Psi\rangle = E | \Psi \rangle$ is characterized by $N_e$
quasi-momenta $-{ \pi\over a} <k_j\le {\pi \over a}$ describing the
motion of the $N_e$ electrons ($a$ is the lattice constant), and $M$
rapidities, $\Lambda_{\alpha} $, describing the spin ordering. Using
the Bethe Ansatz\cite{Bethe} (BA) and imposing periodic boundary
conditions, it can be shown that the quasi-momenta and the rapidities
satisfy the so-called Bethe Ansatz equations \cite{wu,Ha}
\begin{equation}
\label{3.5}
\exp{(i k_jNa)}= \prod_{\alpha=1}^{M}{ \sin{\left( k_j a\right)}
-\Lambda_{\alpha} +{U \over 4} i \over \sin{\left( k_j a\right)}
-\Lambda_{\alpha} -{U \over 4} i}
\end{equation}
\begin{equation}
\label{3.6}
\prod_{j=1}^{N_e}{ \Lambda_{\alpha}- \sin{\left(k_ja\right)} + {U
\over 4} i \over \Lambda_{\alpha}- \sin{\left(k_ja\right)} - {U \over
4} i } = - \prod_{\beta=1}^{M} { \Lambda_{\alpha}- \Lambda_{\beta}
+{U\over 2} i \over \Lambda_{\alpha}- \Lambda_{\beta} -{U\over 2} i }
\end{equation}
The total energy and the total crystal momentum are then given by
\begin{equation}
\label{3.7}
E = -2 \sum_{j=1}^{N_e} \cos{\left(k_j a\right)} \hspace{20mm}
P=\sum_{j=1}^{N_e} k_j
\end{equation}
The ground state is always given by real $k$'s and
$\Lambda$'s. Excited states are usually described by complex
rapidities in so-called ``string'' structures.\cite{wu}

We solve the Bethe Ansatz equations iteratively, starting with a guess
for the set of real rapidities $\Lambda_{\alpha}$ (related to the
ground-state solution of the 1D Heisenberg chain, as described in
Reference \cite{Ha}). Then, we solve Eq.(\ref{3.5}) and find the
quasi-momenta $k_j$, which we use in Eq.(\ref{3.6}) to find the new
set of real rapidities. The procedure is repeated until
self-consistency is reached.

We can check our numerical procedure in two particular cases. First,
the ground-state energy of a half-filled $N=N_e$ chain in the
thermodynamic limit is known to be given by \cite{wu,Fulde}
\begin{equation}
\label{3.8}
E= -4|t|N \int_{0}^{\infty} {dx J_0(x) J_1(x) \over x\left(1+
\exp{(xU/(2|t|))}\right) }
\end{equation}
where the $J_{\nu}(x)$ are cylindrical Bessel functions. In
Fig. \ref{fig3.1} we plot the ground-state energy per site, in units
of $t$, obtained for a half-filled chain of various lengths $N$, for
$U/t=5$. While for very low values of $N$ there are large variations
between the energies of chains with even and odd number of unit cells,
as $N$ increases the energies obtained converge towards the
thermodynamic value of Eq. (\ref{3.8}) (shown as the full line).  We
consider chains with an even number of sites (integer number of unit
cells), since we know that the ground-state has AFM
correlations. Similar curves are obtained for other values of $U/t$.

Another well known case is that of a chain with just one hole, in the
$U/t \rightarrow \infty$ limit. In this limit, double occupancy is
forbidden by the large on-site interaction, and the only possible
charge fluctuation is the motion of the hole. The total energy of the
chain reduces to the energy of the hole, since the single occupied
sites give little contribution to energy (in this limit the
contributions from the AFM correlations of the electron spins, of the
order of $t^2/U$, are negligible for finite chains). It is
straightforward to show \cite{Mattis} that the hole's dispersion
relation in 1D is exactly that of a free particle, and therefore the
total energy of the chain containing a single hole with momentum
$\hbar k$ is $E(k)=-2t \cos{(ka)}$. In the following sections, we show
that the ground-state energy approaches $E=-2t$ as $U/t \rightarrow
\infty$ for the CI method as well as for the Bethe ansatz equations of
the chain with one hole.

\section{Comparison between the BA, the CI and the HF results}

\subsection{The undoped ground-state}

\subsubsection{Hartree-Fock results}

For the undoped system, the self-consistent HF equations
(\ref{2.5},\ref{2.6},\ref{2.7}) for an infinite system give rise to a
mean-field ground state with long range antiferromagnetic
order. However, the Mermin-Wagner theorem states that the true
ground-state of any one dimensional isolated system cannot have long
range order (LRO) and that LRO is absent in 2D systems for non-zero
temperatures. In the framework of the CI method, mobile solitons in
the AFM background mediate the destruction of LRO.

Using the spin and charge distributions $Q(i)=1$ (one electron per
site) and $\vec{S}(i)=(-1)^iS\vec{e}_z$ (AFM order in the arbitrary
direction $\vec{e}_z$), equation (\ref{2.5}) yields two electronic
bands characterized by the dispersion relations
\begin{equation}
\label{2.10}
E^{c/v}_{k\sigma}= \pm \sqrt{\epsilon_k^2+(US)^2} \hspace{10mm} k\in
 \left(-\pi/ 2a, \pi/ 2a\right]
\end{equation}
where $\epsilon_k=-2t\cos{(ka)}$ is the dispersion relation of
non-interacting electrons and $a$ is the lattice constant.  Each of
these levels is double-degenerate ($\sigma=\pm1$).  Given the symmetry
of the spectrum and the fact that only half the states are occupied,
one can easily see that in the Hartree-Fock ground state, all the
states in the valence band $\left(E^v_k <0\right)$ are occupied, while
all the states in the conduction band $\left(E^c_k>0\right)$ are
empty. The two bands are separated by the usual Mott-Hubbard charge
transfer gap opened at the Fermi surface $\left( k=\pm \pi/2a\right)$,
of magnitude $2US$.

Using the valence band wavefunctions in Eqs. (\ref{2.6}) and
(\ref{2.7}), we obtain \cite{MB2} the self-consistent spin amplitude
\begin{equation}
\label{2.12}
S= {US \over N} \sum_{k}^{}{1 \over \sqrt{\epsilon_k^2+(US)^2}}
\end{equation}
where $N$ is the number of sites and the sum is performed over the
Brillouin zone $k\in \left(-\pi/ 2a, \pi/ 2a\right]$. This equation
has three solutions. One is trivial ($S=0$).  For $S\neq 0$ the
equation depends only on $S^2$. Consequently, {\it the mean-field
ground state is doubly-degenerate}: both $+S$ and $-S$ satisfy
Eq. (\ref{2.12}) and give rise to self-consistent ground-states which
differ from each other only through the fact that all the spins are
flipped from one ground-state to the other one.  Strictly speaking,
the HF solution gives an infinite number of degenerate ground states,
because the direction $\vec{e}_z$ is arbitrary (this is a direct
consequence of the fact that the Hubbard model is rotationally
invariant). However, since a real chain is embedded in a 3D crystal,
crystal field interactions will lift the rotational degeneracy, and
fix one particular direction of orientation (easy axis) for the spins
(for instance, along the chains). Consequently one particular
direction $\vec{e}_z$ is favored, and the mean-field ground state is
doubly degenerate.

Since all the states of the valence band are occupied, the energy of
the HF ground state is simply given by:
\begin{equation}
\label{2.13}
E_{GS}=2\sum_{k}^{} E^v_k +NU\left(S^2+{1\over 4}\right)
\end{equation}
where $S$ is given by the self-consistency condition (\ref{2.12}). The
magnitude of the ground state energy per site, in units of $t$, is
plotted in Fig. (\ref{fig2.2}) as a function of $U/t$ (the full line).
The following features can be observed: in the $U/t \rightarrow 0$
limit (noninteracting electrons), the energy of the ground state has,
indeed, the expected value $ E_{GS} \rightarrow -4Nt/ \pi$. In the
strong interaction limit $U/t \rightarrow\infty$, the energy of the
ground state goes to zero as expected (since in this limit every site
is single occupied and virtual hopping is suppressed). For comparison,
the exact ground-state energy obtained from the Bethe Ansatz
(Eq. (\ref{3.8})) is also shown (dashed line). The asymptotic value of
the HF energy is found to be given by $ E_{GS}\rightarrow -2Nt^2/ U$.
It is well known that in this limit, the Hubbard model is equivalent
to an AFM Heisenberg model,\cite{PWA1} with a coupling constant
$J=4t^2/ U$, and that its true ground state energy per site is
\cite{Bethe} $ E_{GS}\rightarrow -NJ \ln{2}=-2.77 Nt^2/U$.  This
suggests that the Hartree-Fock method is a good starting point, from
which to incorporate fluctuation corrections which lower the energy.

\subsubsection{Configuration Interaction treatment of the undoped
chain}

The HF Slater determinant of the undoped AFM ground state is invariant
to translations by $2a$ (AFM order must be preserved). While it is
possible to include in the CI set of Slater determinants excited HF
states of the undoped chain obtained by exciting electrons from the
valence to the conduction band, it turns out that lower energy
self-consistent HF configurations can be generated by breaking the
translational symmetry of the undoped mean-field AFM background.  This
is facilitated by considering excited states of the AFM background
which can accommodate charge carriers in localized states deep within
the charge transfer gap rather than within the Mott-Hubbard bands.
The lowest energy self-consistent excited state is the undoped
(neutral) domain-wall, which describes tunneling from one mean-field
ground-state to the other mean-field ground-state.  Since the AFM
order rotates by $\pi$ when crossing the domain wall, we consider
either one domain wall on an odd-site chain, or a pair of domain walls
on an even-site chain, in order to impose cyclic boundary conditions.
Fig.\ref{fig3.10} depicts a typical self-consistent configuration
containing two neutral domain walls, one centered between sites 10 and
11, and one centered between sites 30 and 31. The charge $Q(i)=1$
everywhere. Near the domain wall the self-consistent spin magnitudes
$S_z(i)$ adjust such that each neutral domain wall carries a spin 1/2
(with a projection in the same direction as its core spins),
suggesting that this excitation is a neutral fermion. This is
confirmed from the electronic structure shown in Fig.\ref{fig3.11}.

We obtain self-consistent configurations containing two neutral
domain-walls at all possible distances from each other, either having
opposite orientations (i.e. total chain spin 0) or same orientations
(total chain spin $\pm 1$). Since the AFM ground state has total spin
0 and states with different total spin do not mix, we need only
include in the set of CI Slater determinants ${|\Psi_i\rangle}$ states
of total spin zero, i.e. those having the neutral domain walls
``paired'' (with opposite orientations).  Since all the possible
configurations with two neutral domain-walls have very similar
energies, they must all be included along with the AFM undoped ground
state in making up the variational trial wavefunction.  We must also
include both AFM undoped mean-field ground-states in the CI set.  This
can be easily seen from Fig.\ref{fig3.10}, where half of the chain is
in one AFM ground state, and the other half is in the other (flipped)
ground state. Therefore, this state will have equal overlap with both
AFM ground states, although the AFM ground-states themselves are
orthogonal to each other.  For an $N$-site chain, the CI set contains
a total of 2+N(N-2)/2 Slater determinants, two being the undoped AFM
ground states, and the rest being the N-2 possible states with paired
domain-walls at different distances from each other, each of which can
be translated N/2 times along the chain.

The total ground-state energy found with the HF (circles), CI
 (squares) and BA (diamonds) methods for chains of different lengths
 $N$ for $U/t=5$ and $U/t=50$ are shown in Fig.\ref{fig3.12}.  For all
 three methods the total energy of the chain is proportional to the
 length of the chain. The addition of the configurations with a pair
 of neutral domain walls in the CI method improves the ground-state
 energy considerably. It is obvious, however, that as the length of
 the chain increases configurations with two, three, four and more
 pairs of neutral domain walls should be included in the CI set in
 order to arrive at a perfect agreement with the exact Bethe Ansatz
 solution.  It is interesting to remark that even if only the
 configurations with one pair of neutral domain walls are included,
 the non-degenerate CI ground state is such that $ \langle
 S_z(i)\rangle =0$ for any site $i$ of the chain, although the
 antiferromagnetic correlations remain very strong. This is a
 consequence of the fact that the CI set of Slater determinants
 contains an equal number of states with the spin at the site $i$ up
 and down, so in average each spin expectation value is
 vanishing. Thus, the CI wavefunction is much more successful in
 mimicking the properties of the exact BA ground-state.

\subsection{Charged solitons in the doped ground state: the spin-bag
 and the charged domain-wall}

\subsubsection{Hartree-Fock results}

If we numerically solve the HF equations
(\ref{2.5},\ref{2.6},\ref{2.7}) for a $N$-site chain with $N-1$
electrons, we find three types of charged self-consistent solutions:
the spin-bag (or spin-polaron) (Fig.\ref{fig2.3}), the charged
domain-wall centered on site (COS domain wall) (Fig.\ref{fig2.4}) and
the charged domain-wall centered between sites (CBS domain wall)
(Fig.\ref{fig2.5}). The spin-polaron is created by trapping the hole
in a small ferromagnetic core, which only affects the LR AFM order
locally. The domain walls are topological excitations, since the AFM
order is rotated by $\pi$ as one goes through the domain
wall. Therefore, in order to impose cyclic boundary conditions, we
must consider an odd-site chain (or we may take an even-site chain and
add two holes, leading to the appearance of a domain wall- anti-domain
wall pair). The localization length of the hole decreases as $U/t$
increasing for all three excitations.  The spin and charge at sites
far from the distortion equal the undoped mean-field ground-state
values.

The electronic spectra corresponding to the configurations shown in
Figs. \ref{fig2.3},\ref{fig2.4} and \ref{fig2.5} are shown in
Figs. \ref{fig2.6} and \ref{fig2.7}. First panel in Fig. \ref{fig2.6}
corresponds to the undoped ground state of a $N=40$ sites chain. As
discussed before, the electronic spectrum consists of two bands of
$N=40$ states each. The valence band is completely filled, the
conduction band is completely empty, and there is a large charge
transfer gap between them. Adding one hole on the same $N=40$ sites
chain, and keeping the cyclic boundary conditions, leads to the
appearance of the spin-bag shown in Fig.\ref{fig2.3}. Its electronic
structure is shown in the right panel of Fig.\ref{fig2.6}. There is a
localized level ($n=1$) well below the valence band, the valence band
contains 38 states, there are 3 localized levels deep inside the
Mott-Hubbard charge transfer gap, and finally the conduction band also
has 38 levels. Since there are $N-1=39$ electrons in the system, only
the localized level below the valence band and the valence band states
are occupied. Since the valence band is spin paired (having an even
number of states), this means that the total spin of this excitation
is 1/2, associated with the spin of the electron on the localized
level. The fact that the spin-bag carries a 1/2-spin is also easy to
deduce from Fig.  \ref{fig2.3}, because of the small ferromagnetic
core. Thus, we conclude that the spin-bag is a charged fermion.

The charged domain wall electronic structures are shown in
Fig. \ref{fig2.7}, with the CBS domain wall in the left panel, and the
COS domain wall in the right panel. In this case, we study a chain
with $N-1=40$ electrons and $N=41$ sites, so that we can impose cyclic
boundary conditions again. We can see that in both cases there are 4
localized levels inside the Mott-Hubbard gap, $N-1=40$ occupied states
in the valence band, and 38 states in the conduction band. The
localized levels of the COS domain wall are degenerate. The degeneracy
is lifted for the CBS domain wall, and the upper discrete level is
pushed quite close to the lower edge of the conduction band. However,
it is still a localized level (this is easily checked by plotting its
wavefunction).  In both cases we have a fully spin-paired valence
band, and therefore the total spin of these excitations is zero. Since
they carry the charge of the hole, the domain walls are charged
bosons, in analogy to the charged solitons of
polyacetylene. \cite{polrev,SSH}

In order to establish the relevance of these different charged spin
configurations, we compare their excitation energies (defined with
respect to the undoped (half-filled) ground state) as a function of
$U/t$ in Fig. (\ref{fig2.9}). As we can see, the domain walls are the
low-energy excitations for $U/t <6.5$, while the spin-polarons become
the low-energy excitations for $U/t >6.5$.  As $U/t \rightarrow 0$,
the core size of the domain-walls diverges roughly like $t/U$. As a
result, in this limit the COS and the CBS domain-walls are very
extended objects which become indistinguishable and
degenerate. However, as $U/t$ increases the core becomes more and more
localized, and the CBS domain-wall becomes energetically favorable
relative to the COS domain-wall.

However, this static HFA does not take into account the lowering of
energy of these excitations due to translations along the chain. From
the simple inspection of the spin distributions of the spin-bag and of
the domain wall, we can easily deduce that while a domain wall can
move freely along the chain, the spin-bag is rather immobile.  Moving
the center of the domain wall by one site (by interchanging the hole
with the spin at the right or at the left) necessitates only some
rearrangement of the magnitude of the core spins, while their
orientation is automatically correct. Consequently, the domain wall
lowers its energy by an amount of order $t$ through hopping along the
chain.  However, if a spin-bag moves only by one site, the translated
spin must be flipped (which would require the the total spin of the
spin-bag to likewise flip). In order to conserve its spin, the
spin-bag must tunnel to the second nearest neighbor. This is a second
order process, and consequently the spin-bag lowers its energy only by
an amount of order $t^2/U$ through motion.  As we demonstrate below,
using the Configuration Interaction method, these qualitative
arguments are valid. When soliton dynamics is incorporated, it is the
charged bosonic domain wall that proves to be the relevant charged
excitation of the Hubbard model for all values of $U/t$.

\subsubsection{Configuration Interaction  treatment of the spin-polaron}
 
Consider a spin-polaron on a chain with $2N$ sites (the number of
sites is even so that we can impose cyclic boundary conditions). Using
the CI method we evaluate the kinetic energy of the mobile, charged
spin-bag. As suggested above, we only need to include in the set of
Slater determinants $|\Psi_i\rangle$ configurations translated by an
even number of sites from the initial HF configuration.  Let
$|\Psi_{even}\rangle, |\Psi_{odd}\rangle$ be the HF determinants for
the spin-polaron centered at an even and odd site, respectively, and
let $\hat{S}_z= \sum_{i}^{}\hat{S}_z(i)={1\over2} \sum_{i,\sigma}^{}
\sigma c^{\dagger}_{i\sigma} c_{i\sigma}$ be the total spin operator
in the $z$-direction.  Then, $\hat{S}_z|\Psi_{even}\rangle={1\over2}
|\Psi_{even}\rangle$ while $\hat{S}_z |\Psi_{odd}\rangle=-{1\over 2}
|\Psi_{odd}\rangle$ (or viceversa), and therefore $ \langle
\Psi_{odd}|\Psi_{even}\rangle=0$. Since the Hubbard Hamiltonian
commutes with $\hat{S}_z$, it follows that $\langle \Psi_{odd}|{\cal
H}|\Psi_{even}\rangle=0$. From the CI equation (\ref{3.3}) we conclude
that there is no mixing between states with the spin-polaron on one
sublattice and states with the (opposite spin) spin-polaron on the
other sublattice.  Therefore, on a chain with $2N$ sites we only need
to mix $N$ Slater determinants in order to obtain the spin-polaron
ground-state within the CI method.

If the initial self-consistent HF spin-polaron configuration
$|\Psi_1\rangle$ is composed of the one-particle occupied orbitals
$\phi^{(1)}_p(i,\sigma)$, the one-particle orbitals of the state
$|\Psi_{n+1}\rangle$ translated by $2na$ will simply be chosen as
$\phi^{(n+1)}_p(i,\sigma)=\phi^{(1)}_p(i-2n,\sigma)$ (cyclic boundary
conditions are assumed). The overlap matrices ${\cal O}_{nm}$ and
${\cal H}_{nm}$ are then calculated and the CI matrix equation
(\ref{3.3}) solved. Numerically, the largest amount of time is spent
computing the ${\cal H}_{nm}$ matrix elements. Due to various
symmetries, there are only $N/2$ distinct matrix elements.

Given the structure of the CI equation (\ref{3.3}), we can readily see
that its solutions are of the form
\begin{equation}
\label{3.9}
|\Psi_k\rangle= \sum_{n=1}^{N} e^{(i k n2a )} |\Psi_n\rangle
\end{equation}
where $a$ is the lattice constant and there are $N$ distinct $k$
 values. These values satisfy the periodicity condition
$\exp{(i k N2a)} =1$
since translating any spin-polaron configuration by the total chain
length $2Na$ leaves the configuration unchanged.  Therefore, the
distinct wavevectors are
$k= m \pi/ Na, m=0,1,...,N-1.$
The reduced Brillouin zone $[0,\pi/a)$ (or, symmetrically, the
$(-\pi/2a, \pi/2a]$ interval) is due to the motion of the spin-polaron
on only one sublattice, and corresponds to states with spin-up (for
instance). The band corresponding to the spin-polaron on the other
sublattice will have the same structure, but corresponds to spin-down
states. In the end we recover the typical spin-degenerate band
expected for fermions.

Given the general form of the wavefunction, the dispersion relation of
the spin-polaron follows from the expression
\begin{equation}
\label{3.12}
E(k)= { \langle\Psi_k|{\cal H}|\Psi_k\rangle\over
\langle\Psi_k|\Psi_k\rangle} = { \sum_{n=1}^{N} \exp{(2ika(n-1))}{\cal
H}_{1n} \over \sum_{n=1}^{N} \exp{(2ika(n-1))}{\cal O}_{1n}}
\end{equation}
In deriving the last equation, we used the symmetry properties of the
matrices ${\cal H}_{nm}$ and ${\cal O}_{nm}$, namely that the $(nm)$
matrix element only depends on $n-m$. Strictly speaking, $E(k)$ is the
energy of the whole chain containing the spin-polaron and will
strongly depend on the length of the chain. We extract the dispersion
relation of the spin-polaron from a fit of the form
\begin{equation}
\label{3.13}
E(k)= 2N e_{GS} + E_{pol}(k)
\end{equation}
Here $e_{GS}$ is interpreted as the ground-state energy per site of
the undoped AFM (for a very long chain, most of the sites are
unaffected by the presence of the single spin-polaron).  We define
$E_{pol}(k)$ as the dispersion relation of the spin-polaron itself. In
other words, the energy of the spin-polaron is defined as the
difference between the energy of the chain with the spin-polaron, and
that of an undoped chain.

We plot $E_{pol}(k)$ versus $k$ in Fig. \ref{fig3.2}, for $U/t=5$ and
chains of various lengths. The various curves fall on top of each
other, thus proving that the fit (\ref{3.13}) is legitimate.  Also
shown is the excitation energy of the static hole-doped spin-polaron
$E_{pol}^{HF}$ (the full line), as obtained from the Unrestricted HFA
(also defined with respect to the energy of the undoped chain).
Clearly, translation lowers the total energy of the spin-polaron, with
the most stable state corresponding to $k= \pi/2a$. The total kinetic
energy gained is, however, only of the order of $t^2/U$. This is shown
in Fig. \ref{fig3.3}, where we plot both the kinetic energy gained
$E_{pol}({\pi\over 2a}) - E_{pol}^{HF}$ (circles) and the width of the
spin-polaron band, $E_{pol}({\pi\over 2a})-E_{pol}(0)$ (squares), as a
function of $t^2/U$. The log-log graph is linear with a slope of unity
as expected, since the spin-polaron must tunnel two sites (second
order hopping process) to the next spin-allowed position.  Clearly,
this charged fermionic excitation is relatively immobile.

We conclude that in the large $U/t$ limit the CI correction to the
spin-bag energy is negligible, due to the immobility of this
excitation. As a result, the energy of the spin-bag varies with $U/t$
as shown in Fig.\ref{fig2.9}, for large $U/t$, and it saturates above
$-1.5t$ as $U/t \rightarrow \infty$.  As already discussed, it is
known that in the $U/t \rightarrow \infty$ limit the energy of the
doping hole is $-2t$.  This discrepancy suggests that the spin-bag
does not provide a good description for the charge carrier.

\subsubsection{Configuration Interaction treatment of 
the charged domain-wall}

To investigate an isolated charged domain wall, we consider chains
with an odd number $(2N+1)$ of sites. As shown in Figs. \ref{fig2.4}
and \ref{fig2.5}, there are two types of self-consistent charged
domain-walls, namely the COS (centered on site) domain wall and the
CBS (centered between sites) domain wall. Since the charged domain
walls are bosons ($S_z|\Psi\rangle=0$), there is non-vanishing overlap
between states with the domain-wall centered on different
sublattices. Unlike the fermionic spin-polarons, we must include all
possible translations in the CI Slater determinant set
${|\Psi_i\rangle}$. For a $2N+1$-site chain, there are $2(2N+1)$
distinct COS domain wall configurations, and $2(2N+1)$ distinct CBS
domain wall configurations. The reason for the factor 2 is that
translation of a domain wall by $2N+1$ sites takes it into a domain
wall centered at the same site as in the initial configuration, but
with all spins flipped, due to the $\pi$-difference in the AFM
ordering of the spins on the two sides of the domain wall. A second
translation around the whole chain is necessary in order to regain the
initial configuration. As a result, there is a four-fold increase in
the number of possible configurations for a domain wall, as compared
to a spin-polaron on a chain of almost the same length. We generate
the translations with an even number $2n$ of lattice sites in the same
way as for the spin-polaron,
$\phi^{(2n+1)}_p(i,\sigma)=\phi^{(1)}_p(i-2n,\sigma)$ if $ 0 \le n \le
N$ (first translation around the chain) and
$\phi^{(2n+1)}_p(i,\sigma)=\phi^{(1)}_p(i-2n, - \sigma)$ if $N < n \le
2N$ (second translation around the chain). Here, we remind the reader
that $\phi^{(m)}_p(i,\sigma)$ represents a particular $(p)$ occupied
one-electron orbital of the static Hartree-Fock Slater determinant
$|\Psi_m\rangle$ which places a static magnetic soliton at site $m$.
For translations with an odd number $2n-1$ of sites, the wavefunction
mapping is $\phi^{(2n)}_p(i,\sigma)=\phi^{(1)}_p(i-(2n-1),-\sigma)$ if
$ 1\le n \le N$ (first translation around the chain) and
$\phi^{(2n)}_p(i,\sigma)=\phi^{(1)}_p(i-(2n-1),\sigma)$ if $ N < n \le
2N+1$ (second translation around the chain).

Let us first consider only including one type of domain wall (either
COS, or CBS), in the CI wavefunction. In this case, we can again
conclude that the solutions of Eq. (\ref{3.3}) must be of the form
\begin{equation}
\label{3.15}
|\Psi_k\rangle= \sum_{n=1}^{2(2N+1)} e^{ikna} | \Psi_n\rangle
\end{equation}
where $|\Psi_n\rangle$ is the configuration translated by $n-1$ sites
from the initial HF configuration $|\Psi_1\rangle$. The periodicity
condition is now
$e^{ik2(2N+1)a}=1$
and the allowed values of $k$ are given by
$k=m \pi/ (2N+1)a, m=0,1,...,2(2N-1)-1$.
Clearly, the domain-wall dispersion band is extended over the full
Brillouin zone $[0,2\pi)$ (or the symmetric version $(-\pi,\pi]$).
The dispersion relation is given by
\begin{equation}
\label{3.18}
E(k)={ \langle\Psi_k| {\cal H}| \Psi_k\rangle \over
\langle\Psi_k|\Psi_k\rangle} = { \sum_{n=1}^{2(2N+1)}
\exp{(ik(n-1)a)}{\cal H}_{n1}\over \sum_{n=1}^{2(2N+1)}
\exp{(ik(n-1)a)}{\cal O}_{n1}}
\end{equation}
As in the case of the spin-polaron, we extract the dispersion relation
of the domain-wall from $E(k)$ by subtracting the energy of the
undoped chain
\begin{equation}
\label{3.19}
E_{dw}(k)=E(k)-(2N+1)e_{GS}
\end{equation}

In Fig. \ref{fig3.5} we show the dispersion relations $E_{dw}(k)$
vs. $k$ for both CBS (left panel) and COS (right panel) domain walls
on chains of different length $2N+1$, and $U/t=5$. The excitation
energy of the static configuration (obtained from the unrestricted HF
search) is also shown. Again, various dispersion curves fall on top of
each other, validating the fit of Eq. (\ref{3.19}). Comparing
Fig. \ref{fig3.5} with Fig. \ref{fig3.2}, it is immediately apparent
that the dispersion band of the domain walls is much wider. In fact,
the band of the COS domain wall extends up to $4t$ (not
shown). Comparing the bottom of the dispersion band to the static HF
excitation energy of domain wall (shown as a full line), we see that
translational motion lowers the energy of the domain wall by about $t$
(as opposed to only $0.3t$ when $U/t=5$ for a spin-polaron). While the
bottom of the dispersion band is basically identical for both types of
domain walls, the top is very different. Excited states with energy
$E(k) > 0$ require the incorporation of the excited state
configurations of the single Slater determinant (from static
Hartree-Fock) in the CI set. Clearly these excited state, static
configurations have energies comparable to the moving domain wall at
high energy parts of the dispersion curve.  The most likely candidates
are those configurations in which electrons from the top of the
valence band are excited into the bound discrete levels. If only one
such excitation takes place, the energy of the static configuration is
raised by $\approx U/2$ (the difference between the energy of the
level at the top of the valence band, and that of the first empty
localized level).  By mixing such configurations in the CI Slater
determinant set, we obtain modifications to the upper part of the
dispersion relation, while the bottom remains unchanged. Since we are
interested in the kinetic energy gained by the domain wall through
translation ($E(k) < 0$ region), we will neglect these higher-energy
processes in what follows.

A technical issue that emerges is the effect of mixing both the COS
and the CBS domain wall configurations when calculating the CI
wavefunction. While one might hope for an improvement in the overall
energy for the mobile, charged soliton, this is not the case. The
reason is that each set of configurations by itself generates
basically the same CI wavefunctions $|\Psi_k\rangle$ at the bottom of
the dispersion band rather than linearly independent ones. This can
easily be seen numerically if we analyze the eigenvalues of the
overlap matrix ${\cal O}_{nm}$. Suppose that $\lambda$ is an
eigenvalue of this matrix, and that $(\alpha_i)_{i=1,N}$ is the
corresponding eigenvector (for simplicity, we use $N$ as the dimension
of the overlap matrix).  Defining $|\Psi\rangle= \sum_{i=1}^{N}
\alpha_i | \Psi_i\rangle$ it is straightforward to show that
$\langle\Psi|\Psi\rangle= \lambda$. When we mix both sets of
configurations together, we find many vanishing eigenvalues
$\lambda=0$, which imply $|\Psi\rangle=0$ (numerically, we use the
Singular Value Decomposition technique as a diagnostic for vanishing
eigenvalues). This proves that there are redundant linearly dependent
combinations in the set of Slater determinants $|\Psi_i\rangle$. We
remove these linearly dependent states to find the CI ground-state. In
particular, by mixing COS and CBS domain wall configurations, the
resulting low-energy spectrum is the same as that found by using only
the lower energy CBS domain wall configurations.

The previous analysis gives us the lowest energy of a single hole
(charged domain wall soliton) on the chain that can be obtained within
the CI approximation. In comparing this energy to the one obtained
from the exact Bethe Ansatz solution, it is not appropriate to
directly compare the total chain energies.  The reason, as already
proved, is that there is a large contribution to these energies
proportional to the number of sites in the chain, the proportionality
constant being the undoped ground-state energy per site (see
Eq. (\ref{3.19})). The HFA gives a higher undoped energy per site than
the exact BA energy (see Fig. \ref{fig2.2}), and the CI approach does
not improve it unless we also add states with pairs of uncharged
domain-walls. Our aim is to isolate the energy of the doping hole.
Therefore, we compare $E_{dw}({\pi\over a})$ (the lowest CI energy of
the domain wall itself) with the corresponding doping hole energy
extracted from the Bethe Ansatz. This comparison is shown in
Fig. \ref{fig3.6}. In order to find the doping hole energy from the
Bethe Ansatz, we evaluate the exact ground-state energy of a chain
with $2N+1$ sites and $2N$ electrons ( half $M=N$ of which have
spin-down), for various values of $N$. This set of energies is seen to
be well fitted by an expression of the form $E(N)=(2N+1)e_{GS}+
E_{0}$, where $e_{GS}$ is in excellent agreement with the BA
ground-state value predicted by (\ref{3.8}). As in the CI approach, we
identify $E_{0}$ with the energy of the hole.  The BA energies of the
hole, as a function of $U/t$ are shown as squares in
Fig. \ref{fig3.6}. They indeed go to $-2t$ in the $U/t \rightarrow
\infty$ limit, as expected. In the $U/t \rightarrow 0$ limit, the
energy of the doping hole is expected to go to zero, since in this
$U=0$ limit the system is a metal.

The CI domain-wall energies $E_{dw}=E_{dw}({\pi \over a})$ are shown
as full circles in Figure \ref{fig3.6}.  The agreement with the Bethe
Ansatz energy is striking. For $U/t \ge 5$, the fit (\ref{3.19}) is
excellent and the error bars on the domain wall energies are extremely
small. However, as $U/t \rightarrow 0$, the size of the domain wall
increases significantly (it is around 20 sites for $U/t=2$) and
therefore extremely long chains need to be considered for a good fit.
The two upper lines correspond to the static HF energies obtained for
a self-consistent CBS domain wall (triangle down) and a COS domain
wall (triangle up). The diamonds show the CI results for the
spin-polaron.  Clearly, the translational motion of the domain wall
(included in the CI approach) drastically lowers its overall
energy. The kinetic energy saved is of the order $t$ over most of the
$U/t$ parameter range.

The agreement between the domain-wall energy as calculated in the CI
approach and the exact doping-hole energy as given by the Bethe Ansatz
is quite remarkable, over the whole range of $U/t$ parameters. The
only disagreement appears for $U/t \le 2$, where the domain-walls
become extremely delocalized and the numerical calculations are very
difficult. The CI solution is not exact because the HF description
neglects the presence of additional neutral domain-walls pairs in the
AFM background. While the addition of such pairs improves the accuracy
of the CI method relative to the exact solution, it makes the
calculation more cumbersome, due to the large increase in the number
of possible configurations.  When the contribution of this background
is removed, the CI solution of a single charged domain wall moving
around the chain agrees very well with the exact solution. This
suggests that the renormalization of the charged domain wall energy by
neutral domain wall pairs is relatively small.  The agreement with the
Bethe Ansatz is not limited to the bottom of the charged domain wall
dispersion curve. In the large $U/t$ limit, the domain-wall dispersion
band is indeed given by $E_{dw}(k)= 2t cos(ka)$, as
required. \cite{Mattis} For example, the dispersion relation for
domain-wall corresponding to $U/t=100$ is shown in
Fig. \ref{fig3.8}. At such high $U/t$ values, the typical energy of
the configurations containing electrons excited on the mid-gap levels
is of order $U/2$. They do not influence the lowest domain wall band.
As $U/t$ is decreased, these excited configurations simply modify the
high energy part of the domain-wall dispersion relation.

 In conclusion, the mobile charged bosonic domain wall excitation is
the relevant charged excitation of the 1D chain described by the
Hubbard model for all values $U/t$. Although the static charged
spin-polaron has lower excitation energy than the static domain-wall
for $U/t > 6.5$, when quantum dynamics is taken into consideration (CI
method) the mobile charged domain wall turns out to be the lowest
energy excitation.

\subsection{\label{ssec3.35}Spin-charge separation}

A particularly striking effect can be recaptured in the CI method if
one adds a hole to an even-site chain. In the HFA, this leads to the
appearance of one spin-polaron, since a single charged domain wall is
incompatible with the cyclic boundary conditions involving an even
number of sites. In the CI method, this charged spin-polaron is
unstable to dissociation into a pair of more mobile domain-wall
solitons. In particular the charged spin-bag dissociates into a
charged bosonic domain-wall and a neutral fermionic domain wall. The
translational kinetic energy saved by the domain wall motion more than
offsets the additional exchange energy cost in creating a pair of
solitons from a single spin-polaron.

We demonstrate this spin-charge separation effect using the set of all
 the possible configurations with a pair of a charged and a neutral
 domain walls in the CI method.  Again, since states with different
 spins do not mix, we need only keep configurations in which the
 uncharged domain wall has the same spin, either +1/2, or -1/2. As in
 the undoped case, configurations with a pair of domain walls connect
 the two possible AFM ground states. This leads to a considerable
 change in the background energy (see Fig.\ref{fig3.12}). As a result,
 we define the energy of the doping hole, in this case, with respect
 to the CI energy of an undoped chain with a pair of neutral
 domain-walls. This allows us to properly account for the
 renormalization of the background energy of the chain by the pair of
 domain walls. Physically, we interpret this in the following way. The
 true ground state of the undoped chain has a certain number of pairs
 of neutral domain walls. When the chain is doped, the doping hole is
 bound into one of the already existing neutral domain walls,
 transforming it into a charged bosonic domain wall. All the other
 pairs of neutral domain walls remain largely unaffected.

The results of this analysis are plotted in Fig.\ref{fig3.11b}.  We
 use the Bethe equations to calculate the ground-state energy of
 even-site chains with one doping hole.  The BA excitation energy for
 the doping hole added to an even-site chain is shown as circles.  The
 exact chain energy is found to be well fitted by $E(N)=Ne_{GS}+E_o$,
 with $N$ being the even number of sites of the chain, and $e_{GS}$
 the undoped ground-state energy per site. Since $Ne_{GS}$ is the
 total energy of the undoped chain, we again identify $E_o$ as the
 energy of the doping hole. The energy of a charged domain wall -
 neutral domain wall pair, obtained using the CI method for even-site
 chains, is shown as squares.  Clearly, the CI method recaptures the
 physics as well as the energetics of the BA solution to a high degree
 of accuracy. Even closer agreement can be achieved if configurations
 with more pairs of uncharged domain-walls are included in the CI
 basis set.  This result provides a very clear illustration of the
 spin-charge separation phenomenon known to exist in the 1D Hubbard
 model. Upon doping, the relevant charge excitation is not the
 quasiparticle excitation (spin-polaron) which carries both the spin
 and the charge of the doping hole, but rather a deconfined pair
 excitation consisting of a charged bosonic domain-wall (carrying
 charge but no spin) and a neutral fermionic domain-wall (carrying
 spin but no charge).

Another interesting result of this analysis is that the excitation
 energy needed to add a hole to an even-site chain is different from
 that needed to add a hole to an odd-site chain (see
 Fig.\ref{fig3.11b}, circles and diamonds).  Such a difference would
 be irreconcilable if the excitation was a local quasiparticle, in
 which case different boundary conditions are expected to introduce
 variations of the order ${\cal O}(1/N)$.  However, in the presence of
 the charge-spin separation, there is a simple interpretation. A doped
 chain with an odd $N$ number of sites has an even $N-1$ number of
 electrons. Its ground state has zero spin (all electrons are
 paired). This is well described by a single charged domain-wall. On
 the other hand, the doped chain with an even $N$ number of sites has
 an odd $N-1$ number of electrons. In order to represent the total
 spin 1/2 of the unpaired electron, it is necessary to include both a
 charged domain wall and a neutral spin-1/2 soliton.  We associate the
 difference $ \Delta E$ between the excitation energy of a single hole
 on odd and even-site chains with the energy of the additional neutral
 spin-1/2 soliton. This interpretation may be verified independently
 by evaluating the energy of a single neutral spin-1/2 domain wall on
 an odd-site, undoped chain.  The BA ground-state energy of the
 undoped odd-site chain varies with the odd-$N$ number of sites as
 $E(N)=Ne_{GS}+E_{dw}$, where $e_{GS}$, the undoped ground-state
 energy per site, has the same value as obtained from fits of
 even-site chains. We identify $E_{dw}$ as the energy of the neutral
 spin soliton. The energies $E_{dw}$ and $\Delta E$ are compared in
 Fig.\ref{fig3.10b}. The good agreement between these independent
 measures of energy confirms that the spin soliton is a well defined
 concept even in the presence of doping. The existence of the spin
 soliton accounts for the difference in energy between odd and even
 site chains even in the thermodynamic limit. It facilitates our
 identification of the spin soliton in the Bethe ansatz with the
 neutral spin-1/2 domain wall of the CI approximation.

\section{ Conclusions}
 
In this paper we have demonstrated the validity of the Configuration
Interaction Method as a technique for going beyond mean field theory
in strongly correlated electron systems. The method recaptures the
essential physical features as well as energetics of the exact
solution of the 1D Hubbard chain. The CI method provides a systematic
way to improve and go beyond the Hartree-Fock Approximation, by
incorporating essential quantum dynamics and tunneling effects which
are excluded from the mean-field theory.

We showed that a charged bosonic domain wall can lower its kinetic
energy by about $t$ for all $U/t$ values, while the immobile charged
spin-polaron can only lower its kinetic energy by an energy of the
order of $t^2/U$. As a result, the mobile charged bosonic domain wall
is the low energy charged excitation of the Hubbard chain for all
values of $U/t$, and its excitation energy and dispersion band are in
good agreement with the predictions of the exact Bethe-Ansatz
solution.  We also showed that it is energetically favorable for the
(quasiparticle-like) charged fermionic spin-bag to dissociate into a
charged bosonic domain wall (which carries the charge but no spin) and
a neutral spin-1/2 domain wall (which carries the spin but no
charge). Clearly the CI method recaptures the physics of spin-charge
separation known to exist in the 1D Hubbard model.  It may be fruitful
to perform a more detailed study of the interaction between various
types of domain walls, both charged and uncharged. Various correlation
functions and response functions can also be calculated for detailed
comparison with exact results.

Given the validity and accuracy of the Configuration Interaction
Method for the 1D problem, we believe that it provides a natural route
to describe effects beyond mean-field theory in the doped ground-state
of the 2D Hubbard model and other models of strongly correlated
electrons in higher dimensions. We present this study
elsewhere.\cite{next}

\section*{Acknowledgments}

M.B. acknowledges support from the Ontario Graduate Scholarship
Program and a fellowship from William F. McLean.  This work was
supported in part by the Natural Sciences and Engineering Research
Council of Canada.

\begin{figure}
\centering
\caption{\label{fig3.1} Ground-state energy per site, in units of $t$,
of an undoped Hubbard chain of $N$ sites and $U/t=5$. The full circles
show the values found directly from the Bethe Ansatz Equations
(\ref{3.5}, \ref{3.6}, \ref{3.7}) while the full line shows the
thermodynamic limit given by Eq.  (\ref{3.8}). In the limit of large
$N$ the two values agree.  }
\end{figure}

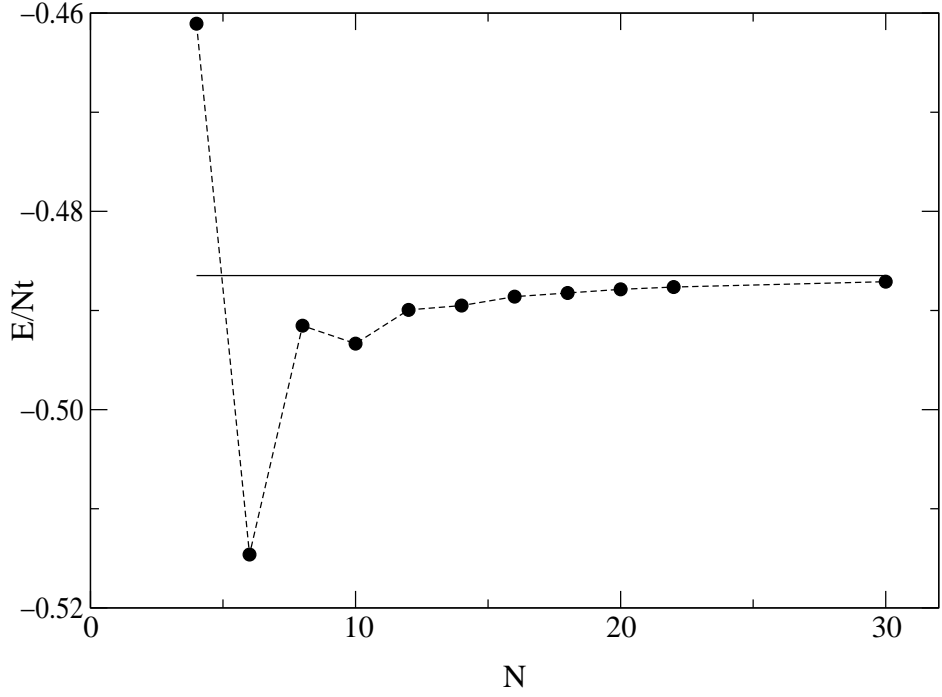
\begin{figure}
\centering
\caption{\label{fig2.2} Energy per site (in units of $t$) of the AFM
undoped background as a function of $U/t$, as obtained from the
Hartree-Fock Approximation (full line) and from the exact Bethe Ansatz
solution (dashed line). }
\end{figure}

\begin{figure} 
\centering
\caption{\label{fig3.10} Self-consistent spin and charge distribution
for a 40-site chain with two neutral domain walls, for U/t=5. The
charge $Q(i)=1$ everywhere. The total spin carried by each neutral
domain wall is 1/2. }
\end{figure}

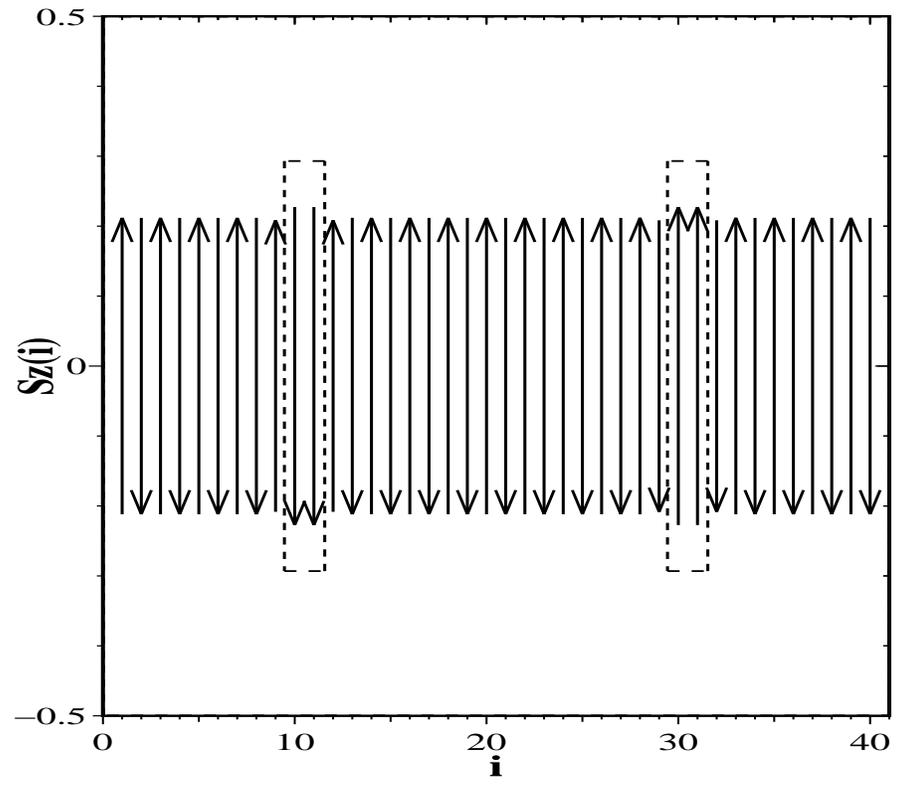
\begin{figure}
\centering
\caption{\label{fig3.11} Electronic structure of the 40-site chain
with the two neutral domain walls shown in Fig.\ref{fig3.10}.  Each
domain wall has 4 discrete levels bound in its core. The spins on the
two occupied bound levels are oriented in the same direction as the
core spins of the domain wall. In the configuration shown in
Fig.\ref{fig3.10} , there is a +1/2 and a -1/2 domain wall, and
therefore all levels are spin paired and the total spin of the chain
is zero. However, for two +1/2 (-1/2) domain walls, all the occupied
discrete levels have +1/2 (-1/2) spins, and the total spin of the
chain is +1 (-1).  }
\end{figure}

\begin{figure}
\centering
\caption{\label{fig3.12} Ground-state energy (in units of $t$) of a
chain of size $N$, calculated with the HF (circles), CI with two
neutral domain walls (squares) and BA (diamonds). The left panel
corresponds to $U/t=5$, while the right one corresponds to
$U/t=50$. Although the energy scale is very different, in both cases
the CI method significantly improves the agreement with the exact
Bethe Ansatz solution.  }
\end{figure}

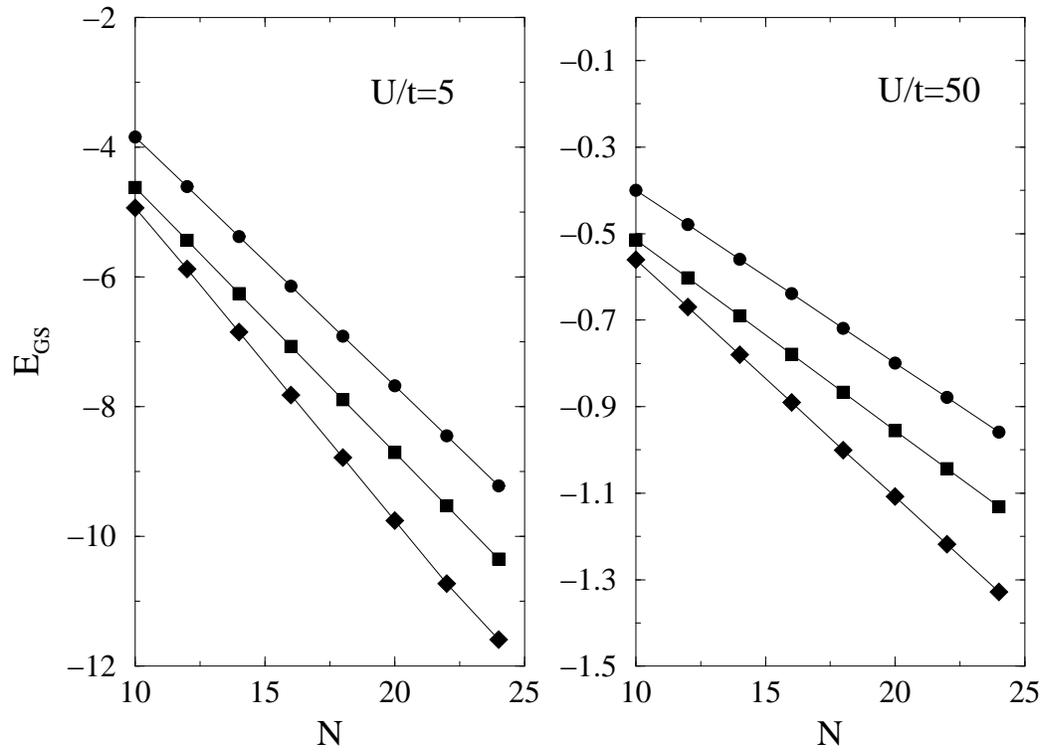
\begin{figure}
\centering
\caption{\label{fig2.3} Self-consistent charge (upper line) and spin
distributions for a charged spin-bag on a 40-site chain, for
$U/t=5$. The spin-bag is a charged fermion. }
\end{figure}

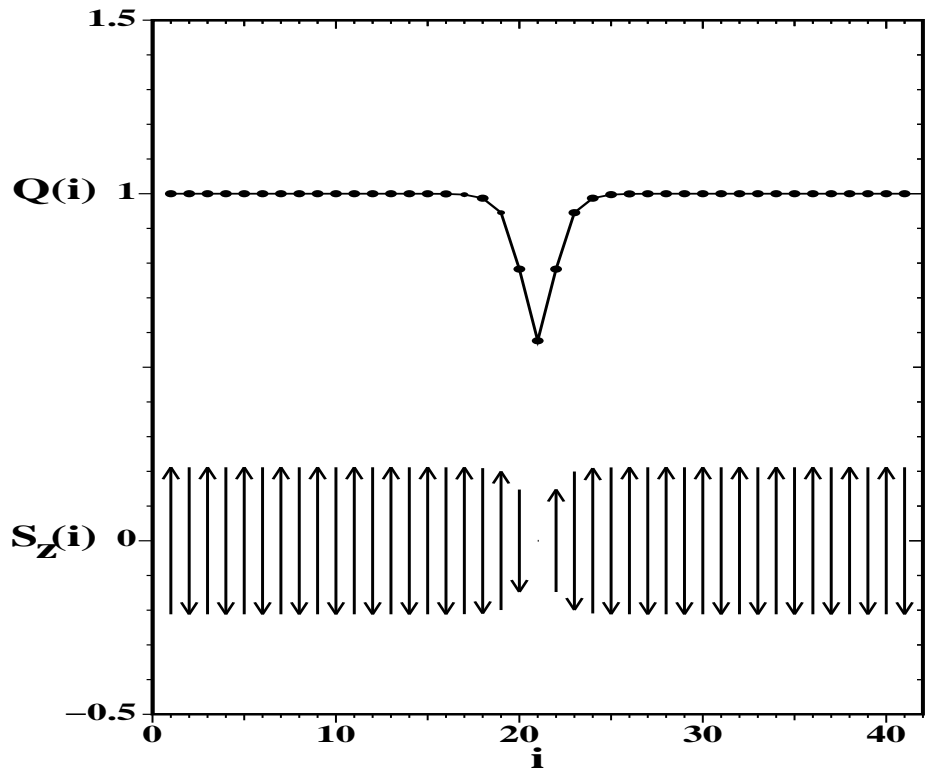
\begin{figure}
\centering
\caption{\label{fig2.4} Self-consistent charge (upper line) and spin
distributions for a domain wall centered on site (COS) on a 41-site
chain, for $U/t=5$.}
\end{figure}

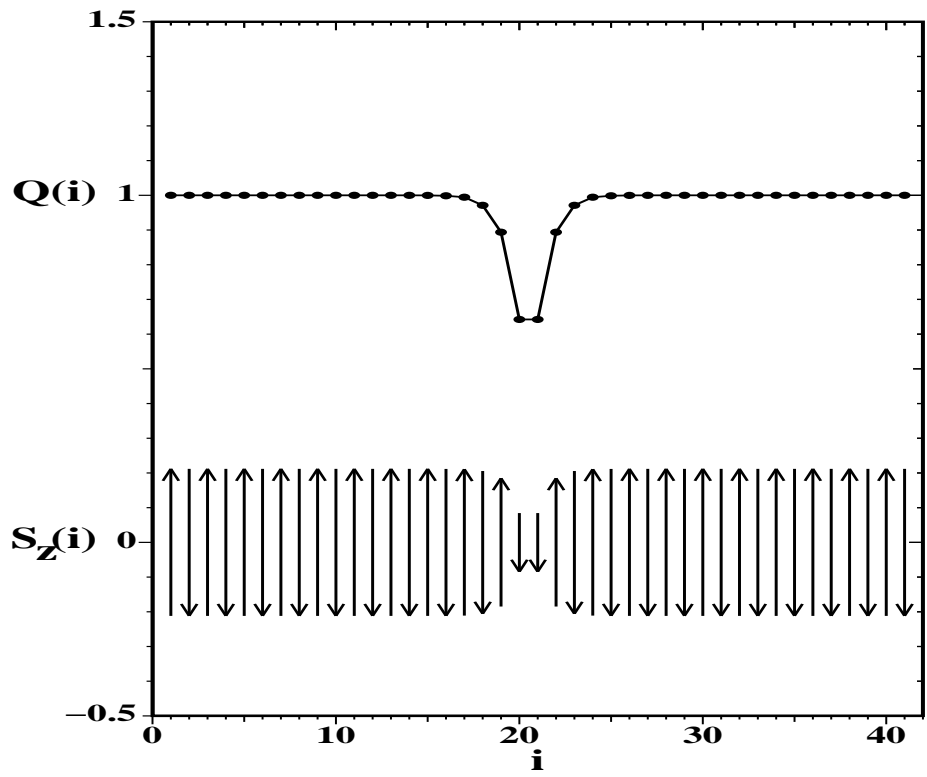
\begin{figure}
\centering
\caption{\label{fig2.5} Self-consistent charge (upper line) and spin
distributions for a domain wall centered between sites (CBS) on a
41-site chain, for $U/t=5$.  The domain walls are charged bosons. }
\end{figure}

\begin{figure}
\centering
\caption{\label{fig2.6} Electronic spectra for an undoped AFM chain
with 40 sites (left panel), and for a 40-site chain with a charged
spin-bag (right panel). $U/t=5$.  }
\end{figure}

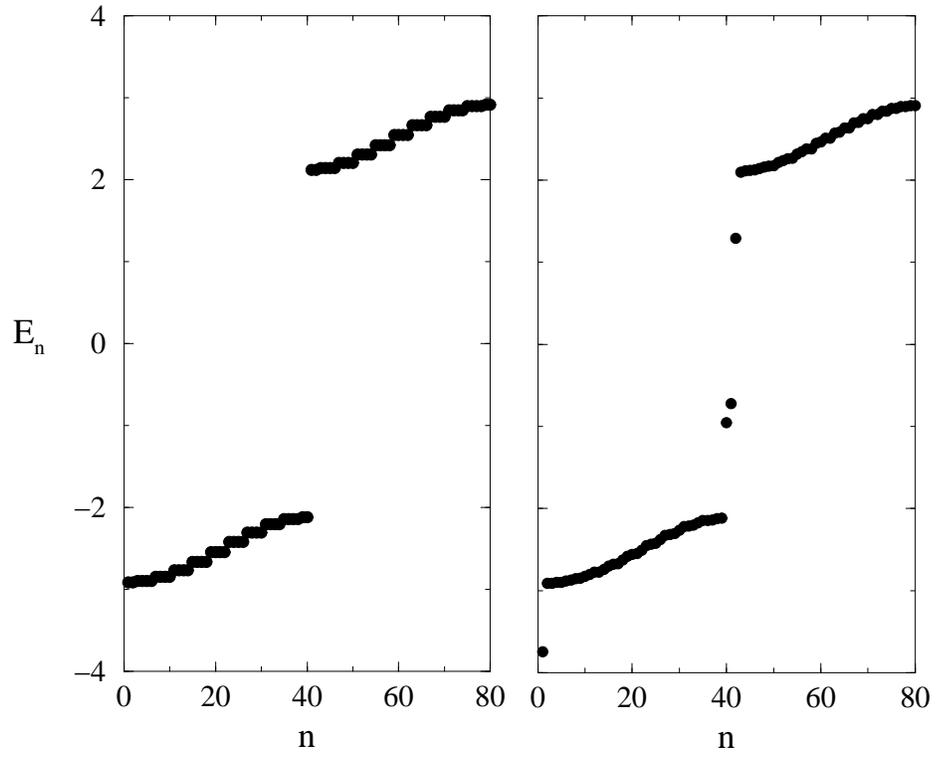
\begin{figure}
\centering
\vspace{10mm}
\caption{\label{fig2.7} Electronic spectra for a chain 41 sites chain
with a CBS domain-wall (left panel), and a COS domain-wall (right
panel). $U/t=5$.  }
\end{figure}

\begin{figure}
\centering
\caption{\label{fig2.9} Excitation energy in units of $t$ of the
charged spin-polaron, CBS and COS domain walls, as a function of
$U/t$. The excitation energy is defined with respect to the undoped
chain. Within the HFA, for $U/t <6.5$ the charged domain-walls are the
low-energy charged excitations, while for $U/t > 6.5$ the
spin-polarons are the low-energy excitations. However, the HFA
approximation neglects the kinetic energy gained by these charged
excitations through translation along the chain. When this is taken
into account within the Configuration Interaction approximation, the
mobile domain-wall is found to be the low-energy charged excitation
for all values of $U/t$ (see Fig. \ref{fig3.6}).  }
\end{figure}

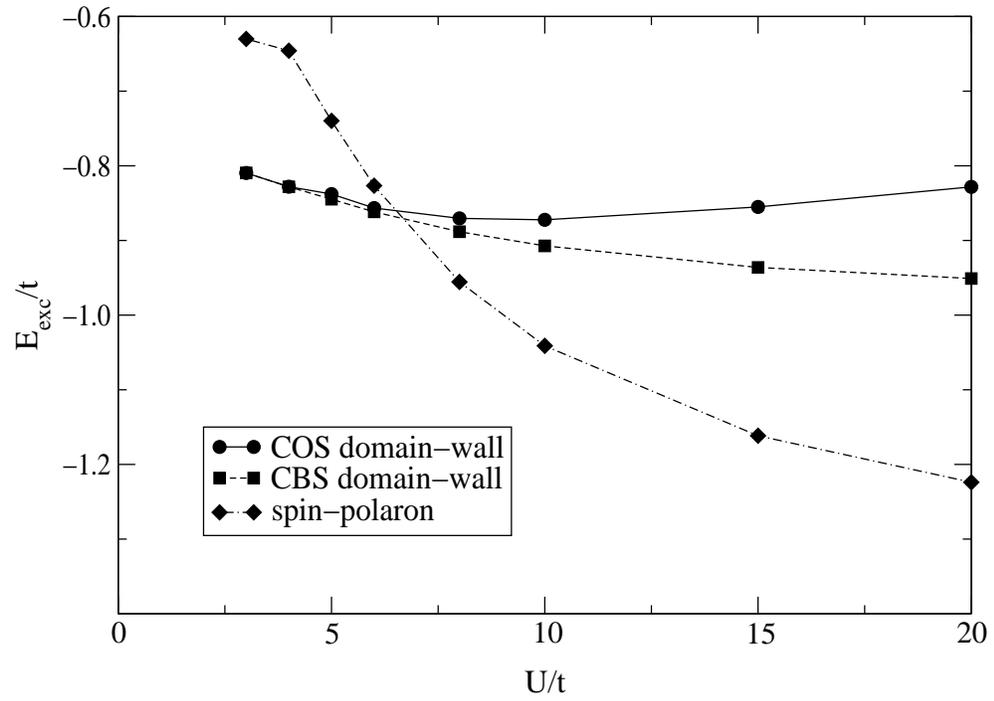
\begin{figure}
\centering
\caption{\label{fig3.2} Dispersion band for the spin-polaron,
$E_{pol}(k)$ vs. $k$, with $E_{pol}(k)$ extracted from
Eq. (\ref{3.13}) for chains of length $2N=14,16, ..., 22$ sites and
$U/t=5$. Also shown is the excitation energy of the static hole-doped
spin-polaron $E_{pol}^{HF}$ (the full line), as obtained from the
Unrestricted HFA.  Translation lowers the total energy of the
spin-polaron, with the most stable state corresponding to $k=
\pi/2a$. However, the kinetic energy gained through translation is
quite small.}
\end{figure}

\begin{figure}
\centering
\caption{\label{fig3.3} The kinetic energy gained by the delocalized
spin-polaron $E_{pol}({\pi\over 2a}) - E_{pol}^{HF}$ (circles) and the
width of the spin-polaron band, $E_{pol}({\pi\over 2a})-E_{pol}(0)$
(squares) as a function of $t^2/U$. The log-log graph is linear with a
slope of unity as expected. The spin-polaron must tunnel two sites to
the next allowed position, through a second order hopping
process. This charged fermion is rather immobile.  }
\end{figure}

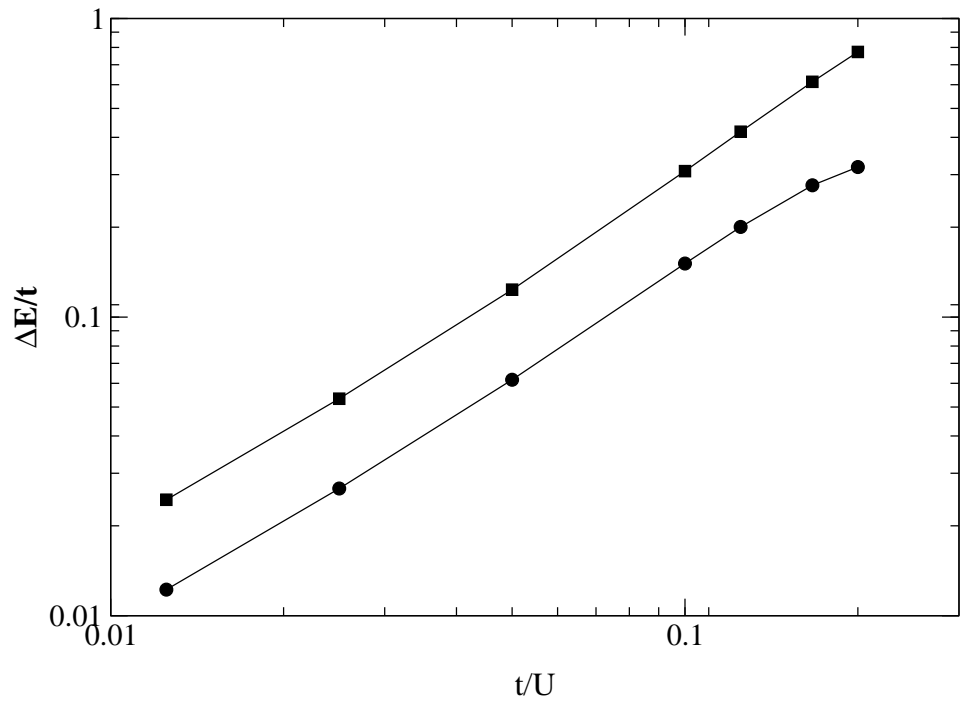
\begin{figure}
\centering
\caption{\label{fig3.5} The dispersion relations $E_{dw}(k)$ vs. $k$
for both CBS (left panel) and COS (right panel) domain walls on chains
of different length $2N+1=17, ..., 23$, and $U/t=5$. The excitation
energy of the static configuration (obtained from the unrestricted HF
search) is also shown as a full line. The extra kinetic energy gained
through translation by the domain-wall is of the order of $t$. }
\end{figure}

\begin{figure}
\centering
\caption{\label{fig3.6} Excitation energy, in units of $t$, for a
mobile charged domain wall (circles) and a mobile charged spin-polaron
(diamonds), as obtained from the CI approach. The exact excitation
energy given by the Bethe-Ansatz method is shown by squares. The
domain-wall CI energy is in excellent agreement with the exact BA
results (also see inset), while the spin-polaron CI energy is
significantly different. For comparison, we also show the excitation
energies for the COS and CBS domain walls as obtained from the static
HFA (up and down triangles), proving again that the extra kinetic
energy gained by the moving domain wall is of order $t$ for most $U/t$
values. In contrast, the extra kinetic energy gained by the
spin-polaron is of order $t^2/U \rightarrow 0$ as $U/t$ increases, so
in the large $U/t$ limit there is almost no difference between the HF
and CI results for the charged spin-polaron. We conclude that the
charged domain-wall is the relevant excitation for all values of
$U/t$.}
\end{figure}

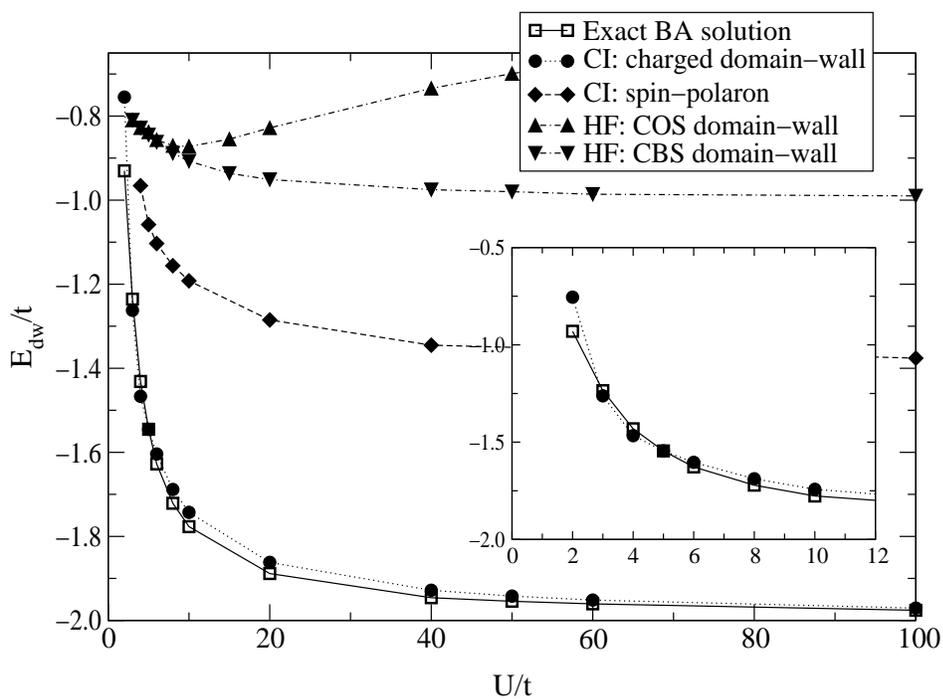
\begin{figure}
\centering
\caption{\label{fig3.8} The dispersion relation for a domain-wall on a
chain of length $2N+1=17,19,21$ and $U/t=100$. In the large $U/t$
limit the dispersion relation of one single hole is given by
$E_{dw}(k)= 2t cos(ka)$. \cite{Mattis} This is indeed in very good
agreement with the dispersion band of the domain wall, proving again
that this is the relevant charged excitation of the Hubbard chain.}
\end{figure}

\begin{figure}
\centering
\caption{\label{fig3.11b} Excitation energy (in units of $t$) for a
charged domain-wall - neutral domain-wall pair, as obtained from the
CI approach for even-site chains (squares). The energy of the doping
hole added to an even-site chain, as obtained from the exact Bethe
Ansatz solution, is shown by circles. Again, there is good agreement
between the two methods. For comparison, we also show the energy of
the doping hole added to an odd-site chain, as obtained from the Bethe
Ansatz (diamonds) and the energy of an isolated charged domain wall on
odd-site chains (triangles). These last two sets are the same as in
Fig.\ref{fig3.6}.}
\end{figure}

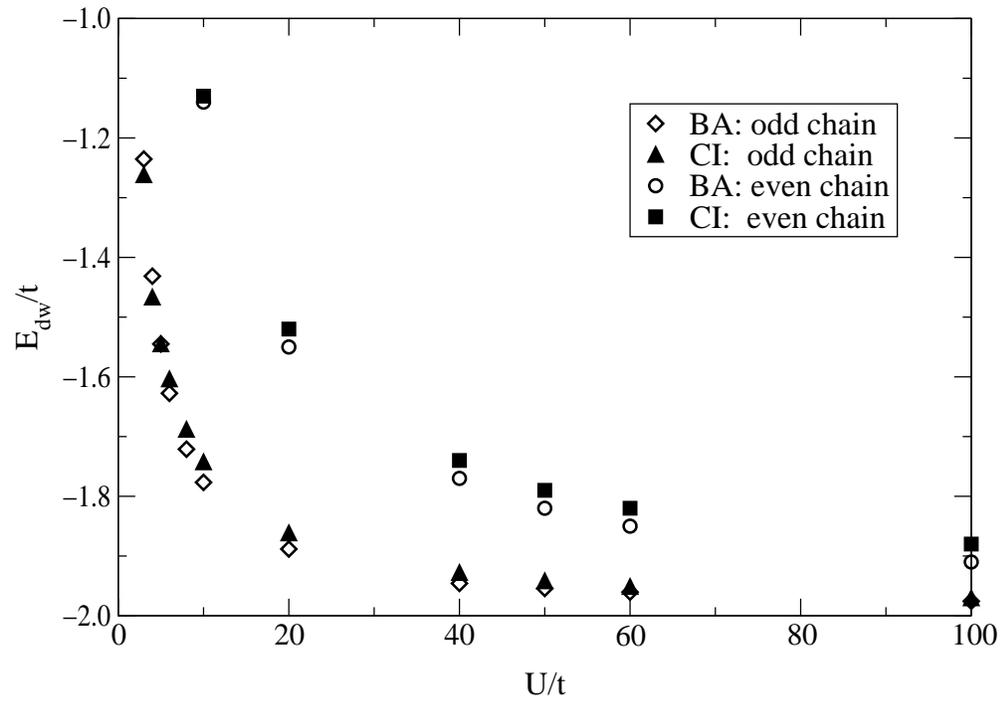
\begin{figure}
\centering
\caption{\label{fig3.10b} The excitation energy of a neutral
domain-wall, as obtained from the Bethe ansatz for undoped odd-site
chains (circles). This agrees with (diamonds) the difference in the
excitation energy of a single hole on an even-site vs.  odd-site chain
(shown individually as circles and diamonds of
Fig.\ref{fig3.11b}). The dotted line is the asymptotic fit $6t^2/U$. }
\end{figure}

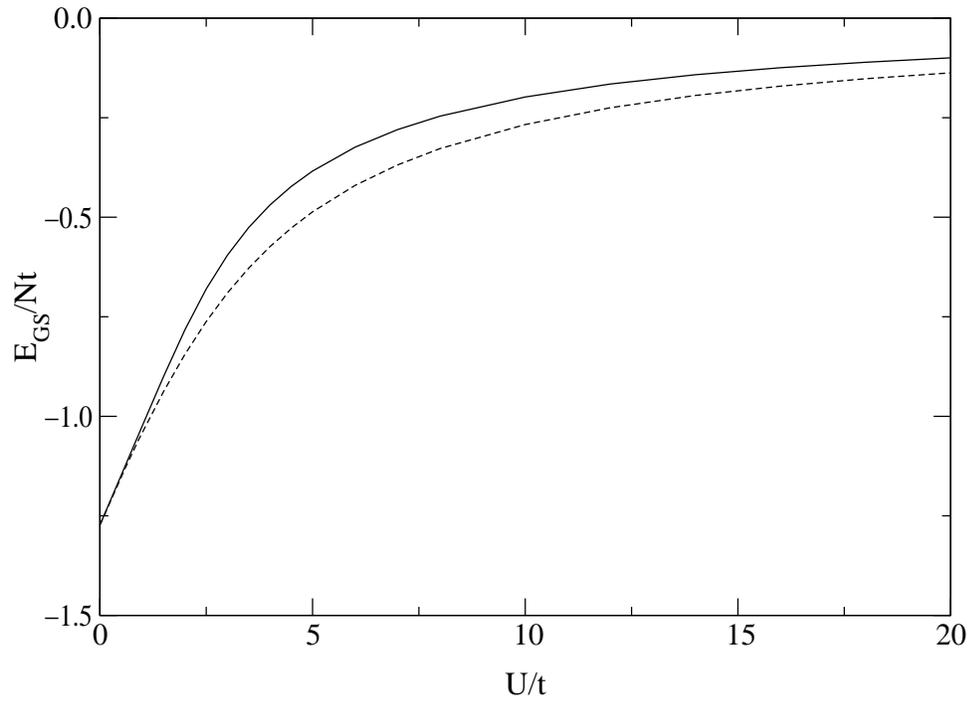
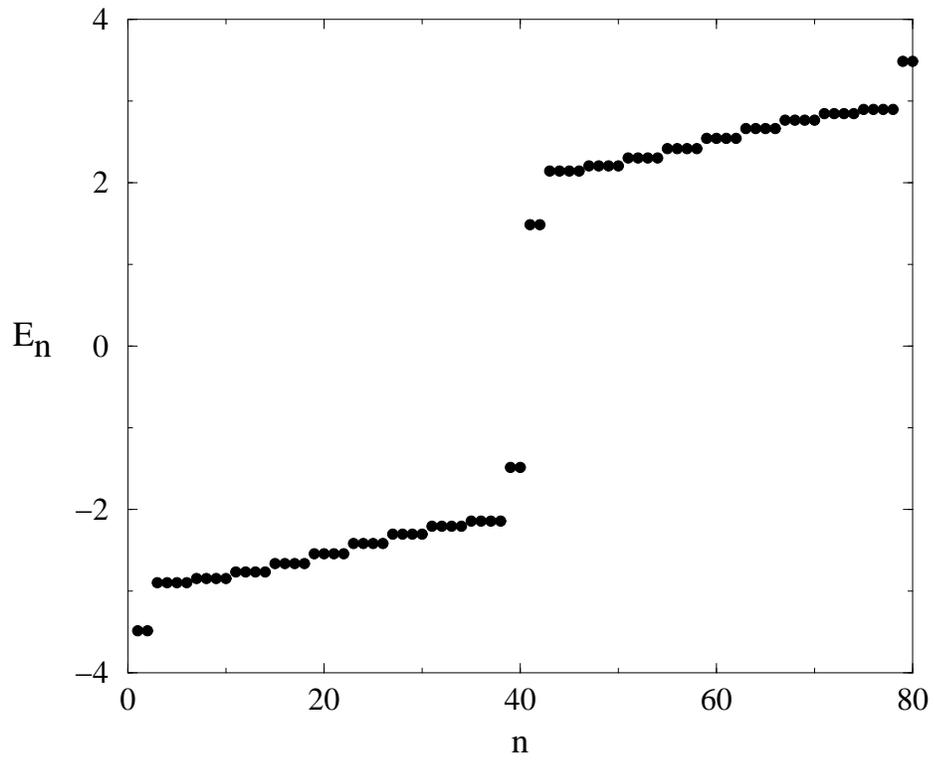
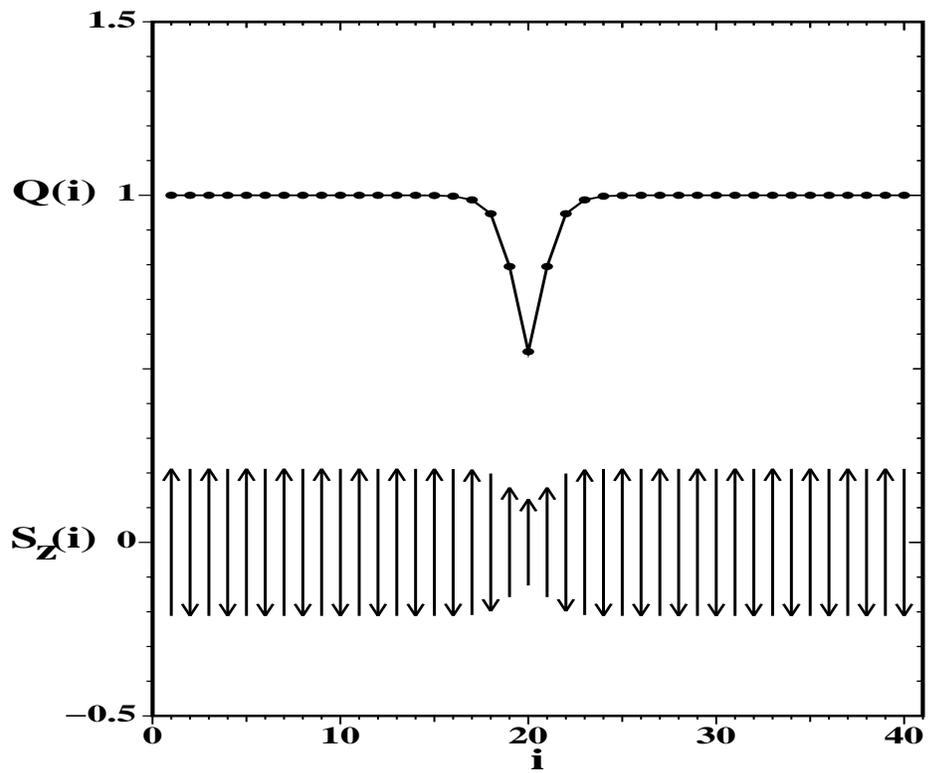
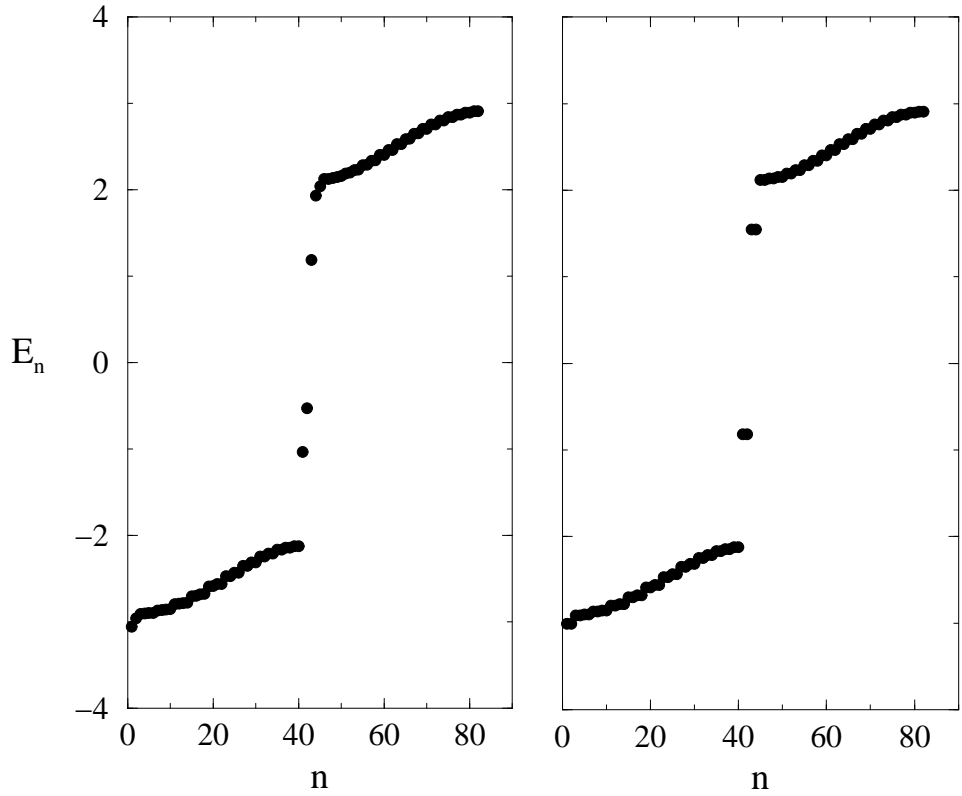
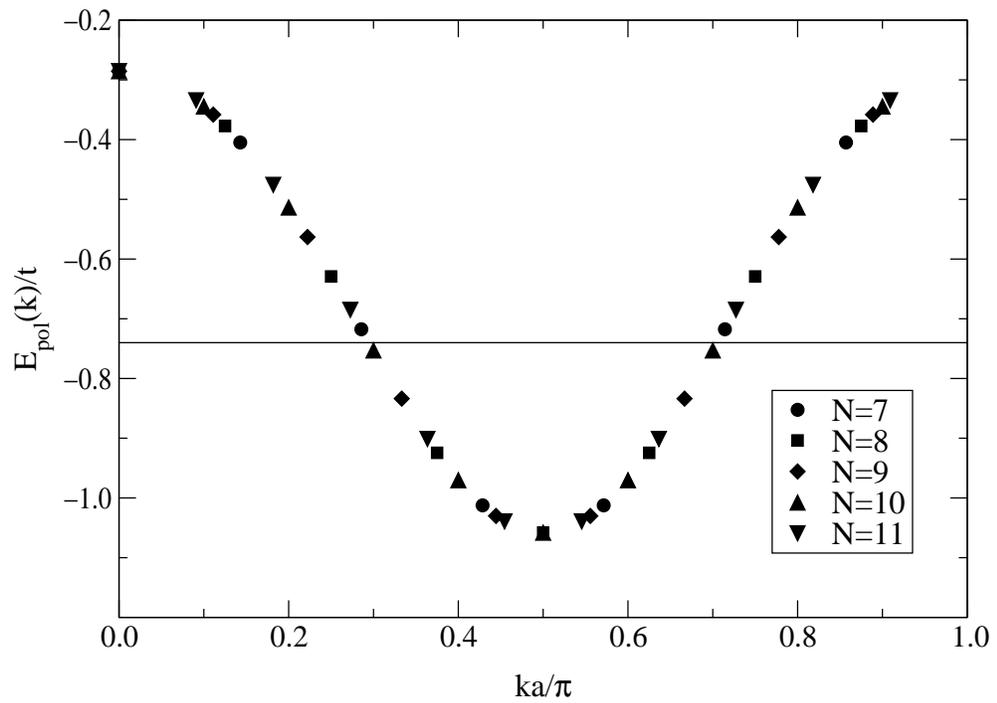
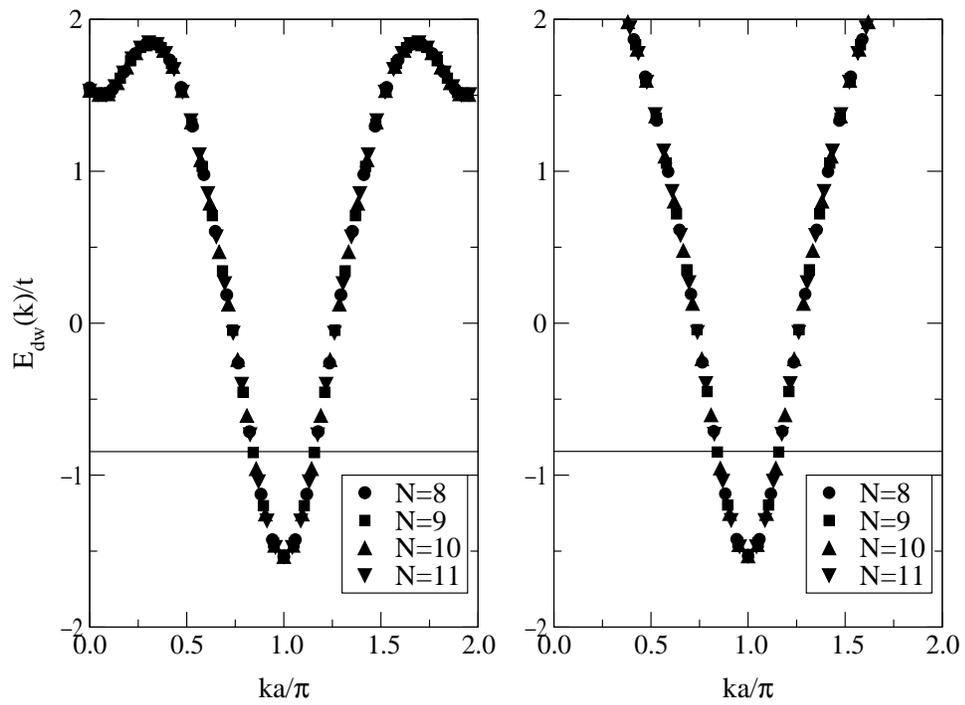
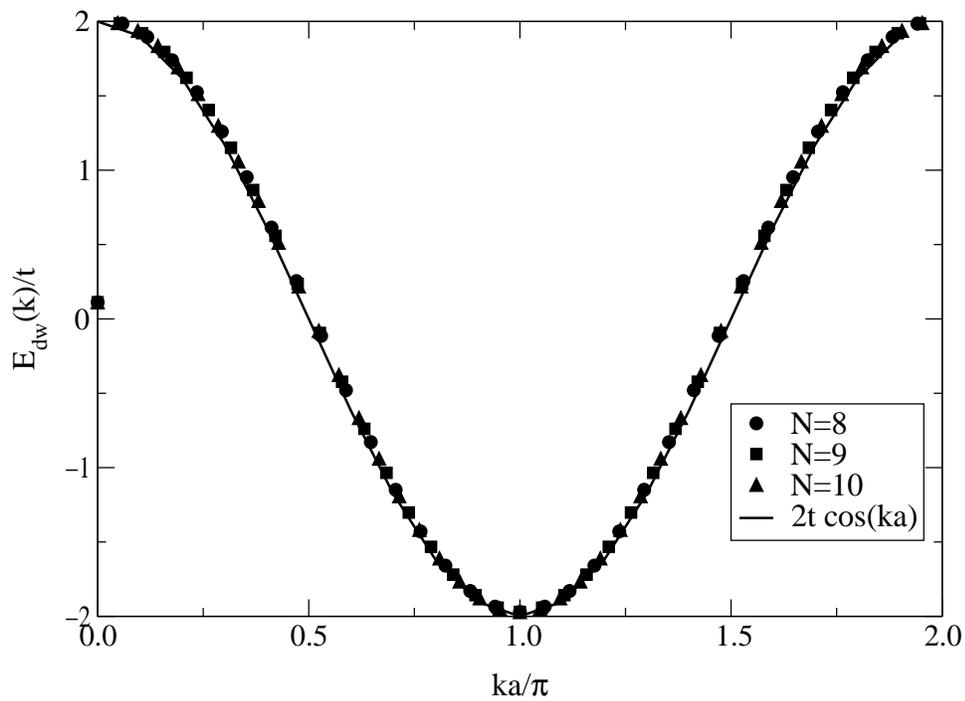
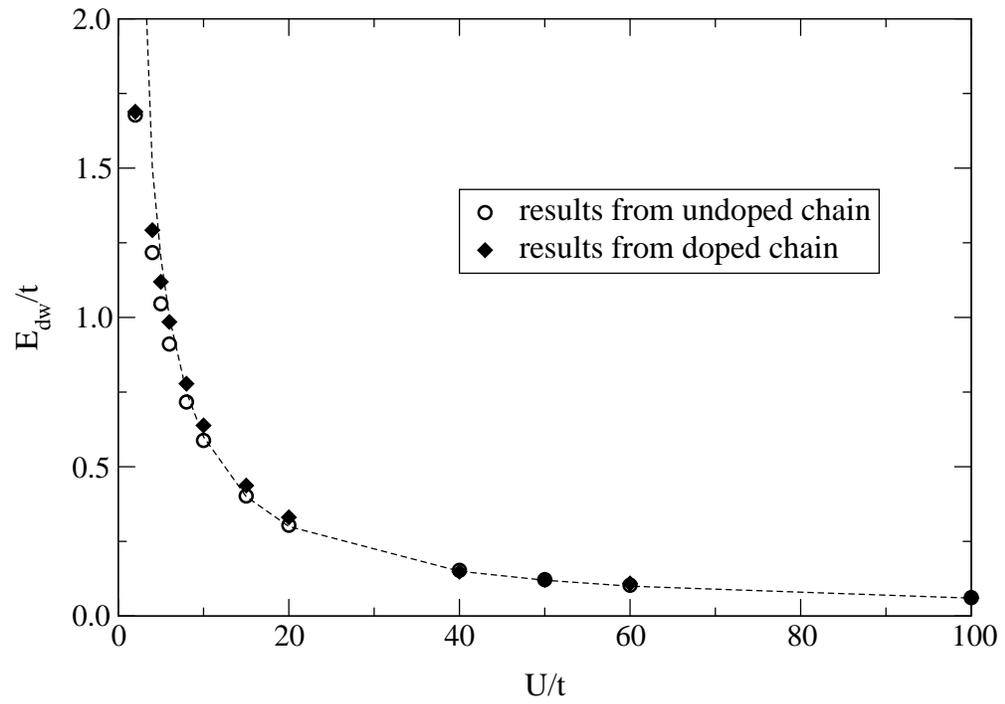
\end{document}